\documentclass[prd,aps,nofootinbib,floatfix,11pt]{revtex4}
\usepackage{amsmath,graphicx,epsfig,amssymb,dsfont,mathtools,}
\usepackage[usenames]{color}
\usepackage{ulem} 
\usepackage{bigstrut}
\usepackage{slashed}
\usepackage{multirow}
\usepackage{subfigure}
\usepackage{color,xcolor}

\allowdisplaybreaks

\begin{document}

\title{Light-Cone Sum Rules Analysis of $\Xi_{QQ^{\prime}q}\to\Sigma_{Q^{\prime}}^{*}$ Weak Decays}
\author{Xiao-Hui Hu$^{1}$~\footnote{Email:huxiaohui@cumt.edu.cn}, Yu-Ji Shi$^{2}$~\footnote{Corresponding author email:shiyuji92@126.com}}
\affiliation{$^{1}$
Cross the Frontier Centre, the College of Materials and physics, China University of mining and technology, Xuzhou 221116, China}
\affiliation{$^{2}$Helmholtz-Institut f\"ur Strahlen- und Kernphysik and Bethe Center for Theoretical Physics,\\ Universit\"at Bonn,
  53115 Bonn, Germany}

\begin{abstract}
  In this work, we investigate the semi-leptonic weak decays of spin-1/2 doubly-heavy baryons $\Xi_{QQ^{\prime}}$ into spin-3/2 singly-heavy baryons $\Sigma_{Q^{\prime}}^*$ within light-cone sum rules. Using the parallel components of the light-cone distribution amplitudes of $\Sigma_{Q^{\prime}}^*$, the transition form factors for these decays are calculated both analytically and numerically. The numerical results for these semi-leptonic weak decays widths and branching ratios are also predicted, which are compared with the same predictions by other theoretical approaches in the literatures. These phenomenology predictions can be tested by the experiments in the future.
\end{abstract}
\maketitle
\section{Introduction}
The doubly-heavy baryons consisting of two heavy quarks (b and c quarks) and one light quark ($u$, $d$, $s$) have been predicted in theoretical literatures a few decades ago~\cite{Gell-Mann:1964ewy,Zweig:1964jf,DeRujula:1975qlm,Jaffe:1975us,Ponce:1978gk,Fleck:1988vm}. As early as 1964, M. Gell-mann and G. Zweig proposed the simple quark model to understand the existence of the numerous observed hadron states \cite{Gell-Mann:1964ewy,Zweig:1964jf}. Furthermore, R. L. Jaffc and J. Kiskis tried to decode the masses of the baryons with the spin-parity $J^{P}=1/2^{+}$ and $J^{P}=3/2^{+}$ by the bag model\cite{Jaffe:1975us}.
For the study of doubly-heavy baryons, a number of comprehensive theoretical researches based on relativistic quark models~\cite{Ebert:2002ig,Roberts:2007ni,Karliner:2014gca,He:2004px,Bagan:1992za}, the Faddeev method~\cite{Valcarce:2008dr}, quantum chromodynamics (QCD) sum rules~\cite{Wang:2010hs,Kiselev:2001fw,Zhang:2008rt,Aliev:2012ru}, potential models~\cite{Richard:2005jz} and lattice QCD~\cite{Lewis:2001iz, Flynn:2003vz,Liu:2009jc} have predicted their masses spectrum, lifetimes and other prospects serving for the detection of doubly heavy baryons.

On the experimental side, the first suspicious signal of  doubly charmed baryon $\Xi_{cc}^{+}$ was claimed by the SELEX collaboration in 2002~\cite{SELEX:2002wqn}. Unfortunately, there is no further evidences from other experiments  to support the claim, which includes the photon-on-fixed-target collisions by the FOCUS collaboration~\cite{Ratti:2003ez} and the $e^+e^-$ collisions by the BaBar~\cite{BaBar:2006bab} and Belle~\cite{Belle:2006edu} collaborations. Perennial research for the doubly heavy baryon, ultimately, the LHC experiment, which is based on the highly integrated luminosity and long-term data accumulation, tells us how the story ends. In 2017, the LHCb collaboration announced the observation of   the doubly charmed baryon $\Xi_{cc}^{++}$ via the decay $\Xi_{cc}^{++}\to \Lambda_{c}^{+}K^{-}\pi^{+}\pi^{+}$~\cite{LHCb:2017iph}. The $\Xi_{cc}^{++}$ mass is measured to be $3620.6\pm1.5(\text{stat})\pm0.4(\text{syst})\pm0.3(\Xi_{c}^{+})\text{MeV}/c^2$. They confirmed the finding by another decay mode $\Xi_{cc}^{++}\to\Xi_c^{+}\pi^{+}$ in 2018~\cite{LHCb:2018pcs}. Now the world averaged value of the $\Xi_{cc}^{++}$ mass is $3621.24\pm0.65(\text{stat})\pm0.31(\text{syst})\text{MeV}/c^2$, which are consistent with the theoretical predictions~\cite{Lewis:2001iz,Karliner:2014gca}.
The lifetime and production cross-section of $\Xi_{cc}^{++}$ is also measured~\cite{LHCb-PAPER-2018-019,LHCb-PAPER-2019-035,LHCb-PAPER-2019-037}.
Nevertheless, the heavier states including the doubly bottomed baryons and charmed-bottomed baryons are still waiting to be found in the future experiments.

To spot all the ground states of the doubly heavy baryons listed in Table.~\ref{tab:JPC} by performing a reconstruction of their weak decays, it is necessary to give reliable theoretical analysis of these decays. Among these weak decays, the semi-leptonic weak decays are suitable for researching the QCD dynamics of doubly heavy baryons theoretically, since all the QCD dynamics can be expressed by the hadron transition matrix elements, which can be decomposed into several form factors.
These transition form factors have been investigated in multifarious methods such as SU(3) flavor sysmetry analysis~\cite{Shi:2017dto,Wang:2021uzi,Li:2021rfj}, QCD sum rules (QCDSR) approach~\cite{Shi:2019hbf,Zhao:2020mod}, light-front quark model (LFQM)~\cite{Wang:2017mqp,Yu:2017zst,Hu:2020mxk}.
Now there is no special approaches which are entirely designed for the doubly heavy baryon physics.
As the techniques are universal, we can also study double heavy baryons via the similar approaches to the one of singly heavy baryons, such as perturbative QCD (pQCD)~\cite{Guo:2005qa,Lu:2009cm}, light cone sum rules (LCSR)~\cite{Wang:2008sm,Wang:2009hra} and so on.
In our previous works, we have applied the LCSR approach for the study of semi-leptonic weak decays of $\Xi_{QQ^{\prime}}\to \Lambda_{Q^{\prime}}$ and $\Xi_{QQ^{\prime}}\to\Sigma_{Q^{\prime}}$, driven by the $c\to d$ and $b\to u$ transition at quark level~\cite{Shi:2019fph,Hu:2019bqj}.
In these semi-leptonic decays, the $\Lambda_{Q^{\prime}}$ and $\Sigma_{Q^{\prime}}$ in the final states are spin $1/2$ and belong to SU(3) antitriplets $\bar{3}$ and sextets $6$ respectively.

The heavy baryon sextets also contain spin $3/2$ states, namely $\Sigma_{Q}^*$, $\Xi_{Q}^{\prime*}$ and $\Omega_{Q}^*$. The aim of this work is to analyze the transition $\Xi_{QQ^{\prime}}\to\Sigma_{Q^{\prime}}^*$ in the LCSR approach.
The non-perturbative dynamics of the quarks and gluons in the baryons are described by the light-cone distribution amplitudes (LCDAs) of $\Sigma_{Q^{\prime}}^*$. The study on these semi-leptonic weak decays can help us to understand the internal strong dynamics within the doubly-heavy baryons. On the other hand, in the future, higher precision measurements on the doubly-heavy baryon decays from various experiments are expected. We hope the future experimental measurements on the $\Xi_{QQ^{\prime}}\to\Sigma_{Q^{\prime}}^*$ transition form factors  can be used to extract more precise heavy baryon LCDAs through the inverse procedure of the LCSR approach performed in our works.

This paper is organized as follows. In Sec.\ref{sec:lc_sum_rules}, the transitions $\Xi_{QQ^{\prime}}\to\Sigma_{Q^{\prime}}^*$ form factors $f_{i}$ and $g_{i}$ are derived in LCSR. In Sec.\ref{sec:numerical results}, we give the numerical results of these form factors, and use them to predict the decay widths and branching ratios of $\Xi_{QQ^{\prime}}\to\Sigma_{Q^{\prime}}^*l\nu_{l}$ decays. In Sec.\ref{sec:conclusions} is a brief summary of this work.

\begin{table*}[!htb]
  \footnotesize
  \caption{Quantum numbers for doubly heavy baryons considered in this paper. The quark content and spin parity $J^{P}$ of these baryons are listed in the second and forth column respectively. The $S_{h}^{\pi}$ is the spin of the heavy quark system.}\label{tab:JPC}
  \begin{center}
  \begin{tabular}{cccc|cccccc} \hline \hline
  Baryon      & Quark Content  &  $S_h^\pi$  &$J^P$   & Baryon & Quark Content &   $S_h^\pi$  &$J^P$   \\ \hline
  $\Xi_{cc}$ & $\{cc\}q$  & $1^+$ & $1/2^+$ &   $\Xi_{bb}$ & $\{bb\}q$  & $1^+$ & $1/2^+$ & \\
  $\Xi_{cc}^*$ & $\{cc\}q$  & $1^+$ & $3/2^+$ &   $\Xi_{bb}^*$ & $\{bb\}q$  & $1^+$ & $3/2^+$ & \\ \hline
  $\Omega_{cc}$ & $\{cc\}s$  & $1^+$ & $1/2^+$ &   $\Omega_{bb}$ & $\{bb\}s$  & $1^+$ & $1/2^+$ & \\
  $\Omega_{cc}^*$ & $\{cc\}s$  & $1^+$ & $3/2^+$ &   $\Omega_{bb}^*$ & $\{bb\}s$  & $1^+$ & $3/2^+$ &  \\ \hline
  $\Xi_{bc}'$ & $\{bc\}q$  & $0^+$ & $1/2^+$ &   $\Omega_{bc}'$ & $\{bc\}s$  & $0^+$ & $1/2^+$ & \\
  $\Xi_{bc}$ & $\{bc\}q$  & $1^+$ & $1/2^+$ &   $\Omega_{bc}$ & $\{bc\}s$  & $1^+$ & $1/2^+$ & \\
  $\Xi_{bc}^*$ & $\{bc\}q$  & $1^+$ & $3/2^+$ &   $\Omega_{bc}^*$ & $\{bc\}s$  & $1^+$ & $3/2^+$ &
   \\ \hline \hline
  \end{tabular}
  \end{center}
  \end{table*}

\section{The LCSR for the Transition $\Xi_{QQ^{\prime}}\to\Sigma_{Q^{\prime}}^*$ Form Factors }
\label{sec:lc_sum_rules}

\subsection{Form Factors}

The transition matrix element of $\Xi_{QQ^{\prime}}\to\Sigma_{Q^{\prime}}^*$ is induced by the $(V-A)^{\mu}$ current, which can be expressed by eight form factors.
$f_{i}$ and $g_{i}$ with $i=1,2,3,4$ are the vector and axial-vector transition currents respectively.
\begin{eqnarray}
  &&	\langle{\Sigma_{Q^{\prime}}^{*}}(P^{\prime},S^{\prime}=\frac{3}{2},S_{z}^{\prime})|(V-A)^{\mu}|{\Xi_{QQ^{\prime}q}}(P,S=\frac{1}{2},S_{z})\rangle \nonumber\\
  && =  \bar{u}_{\alpha}(P^{\prime},S_{z}^{\prime})\Big[{f}_{1}(q^2)\frac{P^{\alpha}}{M_{1}}(\gamma^{\mu}-\frac{\slashed q}{q^2}q^{\mu})+{f}_{2}(q^2)\frac{P^{\alpha}}{M_{1}^2}(\frac{M_{1}^2-M_{2}^{2}}{q^2}q^{\mu}-{\cal P}^{\mu})\nonumber\\
  &&\quad\quad\quad\quad\quad\quad
  +{f}_{3}(q^2)\frac{P^{\alpha}}{M_{1}^2}\frac{M_{1}^2-M_{2}^{2}}{q^2}q^{\mu}+{f}_{4}(q^{2})(g^{\alpha\mu}-\frac{q^{\alpha}q^{\mu}}{q^2})\Big]\gamma_{5}u(P,S_{z})\nonumber\\
  & &\quad- \bar{u}_{\alpha}(P^{\prime},S_{z}^{\prime})\Big[{g}_{1}(q^2)P^{\alpha}(\gamma^{\mu}-\frac{\slashed q}{q^2}q^{\mu})+{g}_{2}(q^2)\frac{P^{\alpha}}{M_{1}^2}(\frac{M_{1}^2-M_{2}^{2}}{q^2}q^{\mu}-{\cal P}^{\mu})\nonumber\\
  &&\quad\quad\quad\quad\quad\quad
  +{g}_{3}(q^2)\frac{P^{\alpha}}{M_{1}^2}\frac{M_{1}^2-M_{2}^{2}}{q^2}q^{\mu}+{g}_{4}(q^{2})(g^{\alpha\mu}-\frac{q^{\alpha}q^{\mu}}{q^2})\Big]u(P,S_{z}),\label{eq:parameterization1}
\end{eqnarray}
here $M_{1}$($M_{2}$) is the mass of $\Xi_{QQ^{\prime}q}$($\Sigma_{Q^{\prime}}^{*}$) and $q^{\mu}({\cal P}^{\mu})=P^{\mu}\mp P^{\prime\mu}$. In the Appendix, we give another parameterization scheme of the above transition matrix element and the relationships between the form factors of them.
In LCSR approach, the following parameterizing scheme is more appropriate to extract the transition form factors compared with the former ones Eq.~(\ref{eq:parameterization1}):
\begin{eqnarray}
  &&	\langle{\Sigma^{*}_{Q^{\prime}}}(P^{\prime},S^{\prime}=\frac{3}{2},S_{z}^{\prime})|(V-A)^{\mu}|{\Xi_{QQ^{\prime}q}}(P,S=\frac{1}{2},S_{z})\rangle \nonumber\\
  && =  \bar{u}_{\alpha}(P^{\prime},S_{z}^{\prime})\Big[\gamma^{\mu}P^{\alpha}{F_{1}(q^{2})}+{F_{2}(q^{2})}P^{\alpha}P^{\mu} +{F_{3}(q^{2})}P^{\alpha}P^{\prime\mu}+F_{4}(q^{2})g^{\alpha\mu}\Big]\gamma_{5}u(P,S_{z})\nonumber\\
  & &\quad-\bar{u}_{\alpha}(P^{\prime},S_{z}^{\prime})\Big[\gamma^{\mu}P^{\alpha}{G_{1}(q^{2})}+{G_{2}(q^{2})}P^{\alpha}P^{\mu} +{G_{3}(q^{2})}P^{\alpha}P^{\prime\mu}+G_{4}(q^{2})g^{\alpha\mu}\Big]u(P,S_{z}),\label{eq:parameterization2}
\end{eqnarray}
The relationships of the form factors ($F_i$ and $G_i$) in Eq.~(\ref{eq:parameterization1}) between the ones defined in Eq. (\ref{eq:parameterization2}) are shown as follows.
\begin{align}
  & {f}_{1}(q^2)=M_{1}F_{1}(q^2),\quad
 {f}_{2}(q^2)=-\frac{M_{1}^2}{2}\Big[F_{2}(q^2)
 +F_{3}(q^2)\Big],\quad
 {f}_{4}(q^2)=F_{4}(q^2),\label{eq:f124}\\
  & {f}_{3}(q^2)=\frac{M_{1}^{2}}{M_{1}^2-M_{2}^{2}}
 \Big[F_{1}(q^2){(-M_{1}-M_{2})}+F_{4}(q^2)\Big]
 +\frac{M_{1}^2}{2}\Big[F_{2}(q^2)
 +F_{3}(q^2)\Big]\nonumber \\
  & \qquad\qquad\quad+\frac{1}{2}\frac{q^{2}M_{1}^2}{M_{1}^2-M_{2}^{2}}
 \Big[F_{2}(q^2)
 -F_{3}(q^2)\Big],\label{eq:f3}\\
  & {g}_{1}(q^2)=M_{1}G_{1}(q^2),\quad
 {g}_{2}(q^2)=-\frac{M_{1}^2}{2}
 \Big[G_{2}(q^2)
 +G_{3}(q^2)\Big],\quad
 {g}_{4}(q^2)=G_{4}(q^2),\label{eq:g124} \\
  & {g}_{3}(q^2)=
 \frac{M_{1}^{2}}{M_{1}^2-M_{2}^{2}}
 \Big[G_{1}(q^2){(M_{1}-M_{2})}+G_{4}(q^2)\Big]
 +\frac{M_{1}^2}{2}\Big[G_{2}(q^2)
 +G_{3}(q^2)\Big]\nonumber \\
  & \qquad\qquad\quad+\frac{1}{2}\frac{q^{2}M_{1}^2}{M_{1}^2-M_{2}^{2}}
 \Big[G_{2}(q^2)
 -G_{3}(q^2)\Big].\label{eq:g3}
\end{align}

\subsection{Light-Cone Sum Rules Framework}	

In order to derive the transition form factors in Eq. (\ref{eq:parameterization2}), we begin from the following correlation function within the framework of LCSR.
\begin{equation}
\Pi_{\mu}(p_{\Sigma^*},q) =i\int d^{4}xe^{iq\cdot x}\langle\Sigma_{Q^{\prime}}^{*}(p_{\Sigma^*})|T\{J^{V-A}_{\mu}(x)\bar{J}_{\Xi_{QQ^{\prime}}}(0)\}|0\rangle\label{eq:corrfunc},
\end{equation}
where $J^{V-A}=\bar{q}_{e}\gamma_{\mu}(1-\gamma_{5})Q_{e}$ is the $V-A$ current, and $J_{\Xi_{QQ^{\prime}}}$ is defined as the interpolating current. In following calculations, we exploit two specific forms of the current $J_{\Xi_{QQ^{\prime}}}$: $J_{\Xi_{QQ}} =\epsilon_{abc}(Q_{a}^{T}C\gamma^{\nu}Q_{b})\gamma_{\nu}\gamma_{5}q^{\prime}_{c}$ with $Q=Q^{\prime}=b$ or $c$, and $J_{\Xi_{bc}} =\frac{1}{\sqrt{2}}\epsilon_{abc}(b_{a}^{T}C\gamma^{\nu}c_{b}
+c_{a}^{T}C\gamma^{\nu}b_{b})\gamma_{\nu}\gamma_{5}q^{\prime}_{c}$. Here $a, b$ and $c$ are the color indices, and $q^{\prime}$ refers the light quark $(u, d, s)$.

We calculate the hadron level  correlation function firstly. It is achieved by inserting a complete set of baryon states between the two currents $J^{V-A}$ and $J_{\Xi_{QQ^{\prime}}}$, in Eq.~(\ref{eq:corrfunc}). In this paper, the contribution of the positive parity states $\Xi_{QQ^{\prime}}^{P+}$ and the negative parity states $\Xi_{QQ^{\prime}}^{P-}$ are considered. Then the correlation function Eq.~(\ref{eq:corrfunc}) at hadron level can be rewritten as
\begin{eqnarray}
  &&\Pi_{\mu}^{hadron}(p_{\Sigma^*},q)=\Pi_{\mu}^{hadron}(p_{\Sigma^*},q)^{+}+\Pi_{\mu}^{hadron}(p_{\Sigma^*},q)^{-}\nonumber\\
  &&\quad=\langle\Sigma_{Q^{\prime}}^{*}(p_{\Sigma^*})|J^{V-A}_{\mu}(0)|\Xi_{QQ^{\prime}}^{+}\rangle\langle\Xi_{QQ^{\prime}}^{+}|\bar{J}_{\Xi_{QQ^{\prime}}}(0)|0\rangle/[m^{2}_{\Xi^{+}}-(p_{\Xi^{+}}-p_{\Sigma^*})^2]\nonumber\\
  &&\qquad+\langle\Sigma_{Q^{\prime}}^{*}(p_{\Sigma^*})|J^{V-A}_{\mu}(0)|\Xi_{QQ^{\prime}}^{-}\rangle\langle\Xi_{QQ^{\prime}}^{-}|\bar{J}_{\Xi_{QQ^{\prime}}}(0)|0\rangle/[m^{2}_{\Xi^{-}}-(p_{\Xi^{-}}-p_{\Sigma^*})^2]+\cdots,\label{correHadron}
  \end{eqnarray}
where the the ellipses denote the contribution of continuum spectra $\rho^{h}$ integration above the threshold $s_{\rm th}$.
The matrix element $\langle\Xi_{QQ^{\prime}}^{\pm}|\bar{J}_{\Xi_{QQ^{\prime}}}(0)|0\rangle$ in Eq.~(\ref{correHadron}), describing doubly heavy baryon $\Xi_{QQ^{\prime}}^{P\pm}$ with the interpolating current $\bar{J}_{\Xi_{QQ^{\prime}}}$
defined by the decay constant $f^{\pm}_{\Xi}$,
\begin{eqnarray}
\langle {\Xi^{P+}_{QQ^{\prime}}}(p_{ \Xi},s)|\bar{J}_{\Xi_{QQ^{\prime}}}(0)|0\rangle&=&f^{+}_{\Xi}\bar{u}_{\Xi}(p_{\Xi},s), \nonumber\\
\langle {\Xi^{P-}_{QQ^{\prime}}}(p_{ \Xi},s)|\bar{J}_{\Xi_{QQ^{\prime}}}(0)|0\rangle&=&-i{\gamma_5}f^{-}_{\Xi}\bar{u}_{\Xi}(p_{\Xi},s).
\end{eqnarray}
Another matrix element $\langle\Sigma_{Q^{\prime}}^{*}(p_{\Sigma^*})|J^{V-A}_{\mu}(x)|\Xi_{QQ^{\prime}}^{\pm}\rangle$ in Eq.~(\ref{correHadron}) has been given by the Eq.~(\ref{eq:parameterization2}). Then the total correlation function at hadron level can be derived. As an example, the one containing only positive parity state contributions and interpolated by the vector current $J_{\mu}^{V}$ can be deduced as
\begin{eqnarray}
  \Pi_{\mu,V}^{hadron}(p_{\Sigma^*},q)^{+}
  &=&-\frac{f^{+}_{\Xi}}{(q+p_{\Sigma^{*}})^{2}-m_{\Xi^{+}}^{2}}
  \bar{u}_{\alpha}(p_{\Sigma^{*}})[F_{1}^{+}(q^{2})P^{\alpha}\gamma^{\mu}+F_{2}^+(q^{2})P^{\alpha}p_{\Sigma^{*}}^{\mu}
  \nonumber\\
  &&\quad+F_{3}^{+}(q^{2})P^{\alpha}p_{\Xi}^{\mu}+F_{4}^{+}(q^{2})g^{\alpha\mu}]\gamma_{5}(\slashed q+\slashed p_{\Sigma^{*}}+m_{\Xi})+\cdots \nonumber\\
  &=&-\frac{f^{+}_{\Xi}}{(q+p_{\Sigma^{*}})^{2}-m_{\Xi}^{2}}\bar{u}_{\alpha}(p_{\Sigma^{*}})\Big[F_{1}^{+}(q^{2})(m_{\Xi}+m_{\Sigma^{*}})q^{\alpha}\gamma^{\mu}+(m_{\Xi}-m_{\Sigma^{*}})F_{3}^{+}(q^{2})q^{\alpha}q^{\mu}\nonumber\\
  &&\quad+[(-m_{\Sigma^{*}}^{2}+m_{\Xi}m_{\Sigma^{*}})(F_{2}^{+}(q^{2})+F_{3}^{+}(q^{2}))-2m_{\Sigma^{*}}F_{1}^{+}(q^{2})]q^{\alpha}v^{\mu}\nonumber\\
  &&\quad+F_{4}(q^2)(m_{\Xi}-m_{\Sigma^{*}})g^{\alpha\mu}-F_{1}^{+}(q^{2})q^{\alpha}\gamma^{\mu}\slashed q-m_{\Sigma^{*}}(F_{2}^{+}(q^{2})+F_{3}^{+}(q^{2}))q^{\alpha}v^{\mu}\slashed q\nonumber\\
  & &\quad-F_{3}^{+}(q^{2})q^{\alpha}q^{\mu}\slashed q-F_{4}(q^2)g^{\alpha\mu}\slashed q\Big]\gamma_{5}+ \cdots.\label{FormextractP}
\end{eqnarray}
The correlation function at hardon level, induced by the current $J_{\mu}^{A}$ and inserted with negative parity states $\Xi_{QQ^{\prime}}^{P-}$ can be calculated via the same procedure.

On the other hand, the QCD level correlation function is calculated by OPE near the light cone $x^2=0$. In the case where the two light quarks form a spin-1 ($j=1$) structure, the heavy baryons can be decomposed into spin-1/2 and spin-3/2 states.
The parallel LCDAs of the baryons with $j=1$ have been given in Ref~\cite{Ali:2012pn}.
\begin{eqnarray}
\frac{\bar{v}^{\mu}}{v_{+}}\langle0|[q_{1}^{T}(t_{1})C\slashed nq_{2}(t_{2})]Q_{\gamma}(0)|H^{j=1}\rangle
&=&\frac{1}{\sqrt{3}}\psi_{\|}^{n}(t_1,t_2)f^{(1)}\epsilon_{\|}^{\mu}u_{\gamma}.\label{eq:pp1}\nonumber\\
\frac{i\bar{v}^{\mu}}{2}\langle0|[q_{1}^{T}(t_{1})C\sigma_{\alpha\beta}q_{2}(t_{2})]Q_{\gamma}(0)\bar{n}^{\alpha}n^{\beta}|H^{j=1}\rangle
&=&\frac{1}{\sqrt{3}}\psi_{\|}^{n\bar{n}}(t_1,t_2)f^{(2)}\epsilon_{\|}^{\mu}u_{\gamma}.\label{eq:pp2}\nonumber\\
{{v}^{\mu}}\langle0|[q_{1}^{T}(t_{1})Cq_{2}(t_{2})]Q_{\gamma}(0)|H^{j=1}\rangle
&=&\frac{1}{\sqrt{3}}\psi_{\|}^{1}(t_1,t_2)f^{(2)}\epsilon_{\|}^{\mu}u_{\gamma}.\label{eq:pp3}\nonumber\\
-{v_{+}}{\bar{v}^{\mu}}\langle0|[q_{1}^{T}(t_{1})C\slashed{\bar{n}}q_{2}(t_{2})]Q_{\gamma}(0)|H^{j=1}\rangle
&=&\frac{1}{\sqrt{3}}\psi_{\|}^{\bar{n}}(t_1,t_2)f^{(1)}\epsilon_{\|}^{\mu}u_{\gamma}.\label{eq:pp4}
\end{eqnarray}
$v$ is the four velocity of the heavy baryon and $n$ is the light-cone vector. Note that $|H^{j=1}\rangle$ is in a reducible representation of the direct-product of spin-1 and spin-1/2. However, what we desire is the irreducible representation of spin-3/2. Therefore, to extract the spin-3/2 component from  $|H^{j=1}\rangle$ one has to conduct a projecting operator on the both sides of each equations in Eq.~(\ref{eq:pp4}). The projecting operator reads as 
\begin{eqnarray}
\left[P_{3/2}\right]_{~\nu\ \gamma}^{\mu~ ~ ~\gamma^{\prime}}=\left[\delta^{\mu}_{\nu}-\frac{1}{3}(\gamma^{\mu}+v^{\mu}\gamma_{\nu})\right]_{\gamma}^{~\gamma^{\prime}},
\end{eqnarray}
where $\gamma,\ \gamma^{\prime}$ are the spinor indexes. Using this operator we can obtain the LCDAs of a purely spin-3/2 heavy baryon:
\begin{eqnarray}
-\frac{2}{3v_{+}}\langle0|[q_{1}^{T}(t_{1})C\slashed nq_{2}(t_{2})]Q(0)|\Sigma_{c}^{*}(v)\rangle
&=&\frac{1}{\sqrt{3}}\psi_{\|}^{n}(t_1,t_2)f^{(1)}u\cdot{\bar{v}}.\label{eq:pp21}\nonumber\\
-\frac{i}{3}\langle0|[q_{1}^{T}(t_{1})C\sigma_{\alpha\beta}q_{2}(t_{2})]Q_{\gamma}(0)\bar{n}^{\alpha}n^{\beta}|\Sigma_{c}^{*}(v)\rangle
&=&\frac{1}{\sqrt{3}}\psi_{\|}^{n\bar{n}}(t_1,t_2)f^{(2)}u\cdot{\bar{v}},\label{eq:pp22}\nonumber\\
-\frac{2}{3}\langle0|[q_{1}^{T}(t_{1})Cq_{2}(t_{2})]Q_{\gamma}(0)|\Sigma_{c}^{*}(v)\rangle
&=&\frac{1}{\sqrt{3}}\psi_{\|}^{1}(t_1,t_2)f^{(2)}u\cdot{\bar{v}},\label{eq:pp23}\nonumber\\
\frac{2v_{+}}{3}\langle0|[q_{1}^{T}(t_{1})C\slashed{\bar{n}}q_{2}(t_{2})]Q_{\gamma}(0)|\Sigma_{c}^{*}(v)\rangle
&=&\frac{1}{\sqrt{3}}\psi_{\|}^{\bar{n}}(t_1,t_2)f^{(1)}u\cdot{\bar{v}}.\label{eq:pp24}
\end{eqnarray}
Here we have also contracted both sides of the equations above by $\bar{v}^{\mu}=\frac{1}{2}\left(\frac{n^{\mu}}{v_{+}}-v_{+} \bar{n}^{\mu}\right)$, and used the fact that $\bar v ^2=-1,\ \bar v\cdot v=0$.
After taking complex conjugate, we can arrive at the following matrix elements 
\begin{eqnarray}
  \epsilon_{abc}\langle\Sigma_{c}^{*}(v)|\bar{q}_{1k}^{a}(t_{1})\bar{q}_{2i}^{b}(t_{2})\bar{Q}_{\gamma}^{c}(0)|0\rangle
  &=&\frac{\sqrt{3}}{16}v_{+}\psi_{\|}^{n*}(t_1,t_2)f^{(1)}\bar{U}_{\gamma}\cdot{\bar{v}}(C^{-1}\slashed{\bar{n}})_{ki}\nonumber\\
  &&-\frac{\sqrt{3}}{16}\psi_{\|}^{n\bar{n}*}(t_1,t_2)f^{(2)}\bar{U}_{\gamma}\cdot{\bar{v}}(C^{-1}i\sigma_{\alpha\beta})_{ki}\bar{n}^{\alpha}n^{\beta}\nonumber\\
  &&+\frac{\sqrt{3}}{8}\psi_{\|}^{1*}(t_1,t_2)f^{(2)}\bar{U}_{\gamma}\cdot{\bar{v}}(C^{-1})_{ki}\nonumber\\
  &&-\frac{\sqrt{3}}{16v_{+}}\psi_{\|}^{\bar{n}*}(t_1,t_2)f^{(1)}\bar{U}_{\gamma}\cdot{\bar{v}}(C^{-1}\slashed{n})_{ki},\label{quarksmatrix1}
\end{eqnarray}
where we have rewritten the spinor notation $u$ as its capital form $U$ to emphasize that now it represents the spinor of the baryon. $a,\ b,\ c$ denote the color indexes. The four parallel LCDAs $\{\psi^{n}_{\parallel}, \psi^{n\bar{n}}_{\parallel},\psi^{1}_{\parallel}, \psi^{\bar{n}}_{\parallel}\}$ are calculated by QCDSR which  correspond to the four LCDAs with different twists $\{\psi_{2}, \psi^{\sigma}_{3},\psi^{s}_{3}, \psi_{4}\}$ in our previous work~\cite{Shi:2019fph}. $f^{(i)}(i=1,2)$ denotes the decay constants of $\Sigma^{*}_{Q}$.

\begin{figure}
\includegraphics[width=0.4\columnwidth]{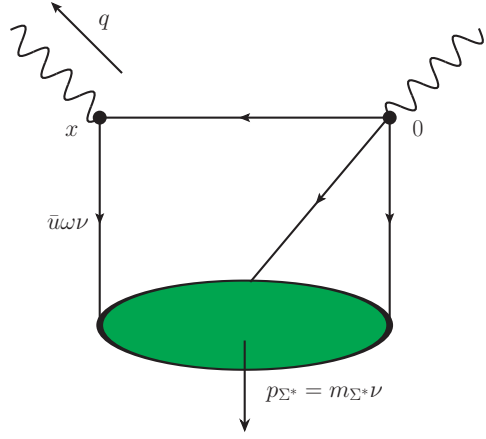}
\caption{Feynman diagram of the QCD level correlation function. The mean of green ellipse is the final state $\Sigma_{Q^{\prime}}^{*}$ with four-velocity $v$. The left black dot represents the $V-A$ current and the right dot indicates the doubly-heavy baryon current. The left straight line describes one light quark in the baryon $\Sigma_{Q^{\prime}}^{*}$. This light quark owns momentum $\bar u \omega v$, where $\bar u$ is its momentum fraction of the diquark momentum.}
\label{fig:FeymDiag}
\end{figure}

Using Eq.~(\ref{quarksmatrix1}), the correlation function induced by vector current $J_{\mu}$ can be given as
\begin{eqnarray}
  \Pi_{\mu}^{QCD}(p_{\Sigma^{*}},q) & =&-\frac{\sqrt{3}i}{8}\int d^{4}x\int_{0}^{\infty}d\omega\omega\int_{0}^{1}due^{i(q+\bar{u}\omega v)\cdot x}\nonumber\\
  &&\times\Big\{v_{+}\psi_{\|}^{n*}(0,x)f^{(1)}\bar{U}\cdot{\bar{v}}(\gamma^{\nu}C(S^{Q}(x))^{T}C^{T}\gamma_{\mu}\slashed{\bar{n}}\gamma_{\nu}\gamma_{5})_{ml}\nonumber\\
  &&-\psi_{\|}^{n\bar{n}*}(0,x)f^{(2)}\bar{U}\cdot{\bar{v}}(\gamma^{\nu}C(S^{Q}(x))^{T}C^{T}\gamma_{\mu}i\sigma_{\alpha\beta}\gamma_{\nu}\gamma_{5})_{ml}\bar{n}^{\alpha}n^{\beta}\nonumber\\
  &&+2\psi_{\|}^{1*}(0,x)f^{(2)}\bar{U}\cdot{\bar{v}}(\gamma^{\nu}C(S^{Q}(x))^{T}C^{T}\gamma_{\mu}\gamma_{\nu}\gamma_{5})_{ml}\nonumber\\
  &&-\frac{1}{v_{+}}\psi_{\|}^{\bar{n}*}(0,x)f^{(1)}\bar{U}\cdot{\bar{v}}(\gamma^{\nu}C(S^{Q}(x))^{T}C^{T}\gamma_{\mu}\slashed{n}\gamma_{\nu}\gamma_{5})_{ml}\Big\}.\label{Corre1}
\end{eqnarray}
Here the light-cone vectors $n, \bar n$ can be expressed as the following Lorentz covariant forms
\begin{equation}
  n_{\mu} =\frac{1}{v\cdot x}x_{\mu},\ \ \ \bar{n}_{\mu}=2v_{\mu}-\frac{1}{v\cdot x}x_{\mu}.\label{nnbar}
  \end{equation}
The heavy quark propagator $S^{Q}(x)$ is given by the following equation,
\begin{equation}
S^{Q}(x) =\int\frac{d^4k}{(2\pi)^4}e^{-ik\cdot x}\frac{(\slashed k-m_{Q})}{k^2-m_{Q}^2+i\epsilon}.\label{propagator}
\end{equation}
During the above derivation, the light-cone components of the two light quark momenta in $\Sigma_{Q}^{*}$ are denoted as $\omega_{1}$, $\omega_{2}$ and $\omega=\omega_{1}+\omega_{2}$. The momenta fraction $u$ is given by $\omega_{2}=(1-u)\omega=\bar{u}\omega$ shown in Fig~\ref{fig:FeymDiag}.
Then the correlation function induced by $J_{\mu}^{V}$ transition current will be expressed as a convolution form of $\omega$ and $u$
\begin{eqnarray}
  \Pi_{\mu,V}^{QCD}(p_{\Sigma^*},q)
  &=&-\frac{\sqrt{3}}{8}\int d^4x\int^{2s_{0}}_{0}d\omega \omega \int^{1}_{0}du\int\frac{d^4k}{(2\pi)^4}e^{i(q+\bar{u}\omega v-k)\cdot x}\nonumber\\
  &&\times\Big\{\psi_{\|}^{n*}(\omega,u)f^{(1)}\bar{u}_{\lambda}[(\frac{x^{\lambda}}{v\cdot x}-v^{\lambda})\gamma^{\nu}\frac{(\slashed k-m_{Q})}{k^2-m_{Q}^2+i\epsilon}\gamma_{\mu}(2\slashed{v}-\frac{\slashed x}{v\cdot x})\gamma_{\nu}\gamma_{5}]_{ml}\nonumber\\
  &&-\psi_{\|}^{n\bar{n}*}(\omega,u)f^{(2)}\bar{u}_{\lambda}[(\frac{x^{\lambda}}{v\cdot x}-v^{\lambda})\gamma^{\nu}\frac{(\slashed k-m_{Q})}{k^2-m_{Q}^2+i\epsilon}\gamma_{\mu}i\sigma_{\alpha\beta}(2{v}^{\alpha}-\frac{ x^{\alpha}}{v\cdot x})\frac{ x^{\beta}}{v\cdot x}\gamma_{\nu}\gamma_{5}]_{ml}\nonumber\\
  &&+2\psi_{\|}^{1*}(\omega,u)f^{(2)}\bar{u}_{\lambda}[(\frac{x^{\lambda}}{v\cdot x}-v^{\lambda})\gamma^{\nu}\frac{(\slashed k-m_{Q})}{k^2-m_{Q}^2+i\epsilon}\gamma_{\mu}\gamma_{\nu}\gamma_{5}]_{ml}\nonumber\\
  &&-\psi_{\|}^{\bar{n}*}(\omega,u)f^{(1)}\bar{u}_{\lambda}[(\frac{x^{\lambda}}{v\cdot x}-v^{\lambda})\gamma^{\nu}\frac{(\slashed k-m_{Q})}{k^2-m_{Q}^2+i\epsilon}\gamma_{\mu}\frac{\slashed x}{v\cdot x}\gamma_{\nu}\gamma_{5}]_{ml}\Big\}\nonumber
\end{eqnarray}
Integrating the space-time coordinate $x$, we obtain the correlation function of $J_{\mu}^{V}$ as
\begin{eqnarray}
&&\Pi_{\mu,V}^{QCD}((p_{\Sigma^*}+q)^2,q^2)\nonumber\\
 &=&\int_{0}^{2s_0} d \omega \int_{0}^{1} d u f^{(1)}\Big\{\sqrt{3} \bar{u}^{2} \hat{\psi}_{ \|}^{n *}(\omega, u) \Big[-2\bar{u}_{\lambda}q^{\lambda}[2(\slashed q+\bar{u}\omega+m_{Q})(q_{\mu}+\bar{u}\omega v_{\mu}) \nonumber\\
 & &\qquad-(q^2+2\bar{u}\omega q\cdot v+\bar{u}^2\omega^2)\gamma_{\mu}]\frac{1}{\Delta^{3}}+\bar{u}_{\lambda}[-2q^{\lambda}\gamma_{\mu}+(\slashed q+\bar{u}\omega+m_{Q})g^{\lambda}_{\mu}]\frac{1}{\Delta^{2}} \Big]\nonumber\\
& &+\sqrt{3}\bar{u} \tilde{\psi}_{ \|}^{n *}(\omega, u)\Big[\bar{u}_{\lambda}q^{\lambda}[\gamma_{\mu}\slashed q-\bar{u} \omega\gamma_{\mu}+2(\bar{u} \omega+m_{Q})v_{\mu}]\frac{1}{\Delta^{2}} -\bar{u}_{\lambda}g^{\lambda}_{\mu}\frac{1}{\Delta} \Big]\Big\}\nonumber\\
 &+&\int_{0}^{2s_0} d \omega \int_{0}^{1} d u f^{(2)}\Big\{\sqrt{3}  \bar{u}^{2} \hat{\psi}_{ \|}^{n \bar{n} *}(\omega, u) \Big[-4\bar{u}_{\lambda}q^{\lambda}[-2(\bar{u} \omega+m_{Q}+q\cdot v)q_{\mu}\nonumber\\
 &&+2(\bar{u}\omega q\cdot v+q^2)v_{\mu}+m_{Q}(\gamma_{\mu}\slashed q+\gamma_{\mu}q\cdot v)]\frac{1}{\Delta^{3}} +2\bar{u}_{\lambda}[5q^{\lambda}v_{\mu}-g^{\lambda}_{\mu}(q\cdot v+\bar{u}\omega)]\frac{1}{\Delta^{2}}
 \Big] \Big\}\nonumber\\
 &+&\int_{0}^{2s_0} d \omega \int_{0}^{1} d u f^{(2)} \Big\{\sqrt{3}\bar{u}\tilde{\psi}_{1}^{1 *}(\omega, u)\Big[\bar{u}_{\lambda}q^{\lambda}[2(q_{\mu}+\bar{u}\omega v_{\mu})+m_{Q}\gamma_{\mu}]\frac{1}{\Delta^{2}}-\bar{u}_{\lambda}g^{\lambda}_{\mu}\frac{1}{\Delta} \Big] \Big\}\nonumber\\
 &+&\int_{0}^{2s_0} d \omega \int_{0}^{1} d u f^{(1)}\Big\{\bar{u}^{2} \hat{\psi}_{ \|}^{\bar{n} *}(\omega, u) \Big[2\bar{u}_{\lambda}q^{\lambda}[\gamma_{\mu}(q^2+2\bar{u}\omega q\cdot v+\bar{u}^2\omega^2)\nonumber\\
 &&-2(\slashed q+\bar{u}\omega-m_{Q})(q_{\mu}+\bar{u}\omega v_{\mu})]\frac{1}{\Delta^{3}}
 +\bar{u}_{\lambda}[g^{\lambda}_{\mu}(\slashed q+\bar{u} \omega-m_{Q})-2q^{\lambda}\gamma_{\mu}]\frac{1}{\Delta^{2}}\Big]\Big\}
 ,\label{correQCD}
\end{eqnarray}
where $\Delta=(q+\bar{u} \omega v)^{2}-m_{Q}^{2}$ and $m_Q$ is the mass of the decaying heavy quark $Q=b ,c$. Note that later we have to extract the discontinuity of the correlation function in terms of the momentum square of the inserted doubly heavy baryon state, namely $(q+\bar{u} \omega v)^{2}$. Such discontinuity only comes from the $\Delta$s in the denominators, and we have to explicitly express  $\Delta$ as a function of $(q+\bar{u} \omega v)^{2}$, which reads as
\begin{eqnarray}
&&\Delta=\frac{\bar{u}\omega}{m_{\Sigma^*}}s
+H(u,\omega,q^{2})-m_{Q}^2,\nonumber\\
&&s=(p_{\Sigma^*}+q)^2,\ \
H(u,\omega,q^{2})=\bar{u}\omega(\bar{u}\omega-m_{\Sigma^*})+(1-\frac{\bar{u}\omega}{m_{\Sigma^*}})q^{2}.
\end{eqnarray}
In Eq.~(\ref{correQCD}), we have introduced two kinds of newly defined LCDAs
\begin{eqnarray}
\tilde{\psi}_i(\omega,u) =\int_{0}^{\omega}d\tau\tau\psi_i(\tau,u),\ \ \
 \hat{\psi}_i(\omega, u) =\int_{0}^{\omega} d \tau \tilde{\psi}_i(\tau, u) \ \ \ \ (i=2,\ 3\sigma,\ 3s,\ 4).\nonumber
\end{eqnarray}

As a consequence, one obtains the correlation function at QCD level, which can be decomposed into eight Lorentz structures $\{q^{\lambda}\gamma_{\mu}, q^{\lambda}v_{\mu}, q^{\lambda}q_{\mu}, g^{\lambda}_{\mu}, q^{\lambda}\gamma_{\mu}\slashed q, q^{\lambda}v_{\mu}\slashed q, q^{\lambda}q_{\mu}\slashed q, g^{\lambda}_{\mu}\slashed q\}$:
\begin{eqnarray}
\Pi_{\mu,V}^{QCD}((p_{\Sigma^{*}}+q)^2,q^2)&=&\bar{u}_{\lambda}(p_{\Sigma^{*}})\{C_{q^{\lambda}\gamma_{\mu}}q^{\lambda}\gamma_{\mu}+C_{q^{\lambda}v_{\mu}}q^{\lambda}v_{\mu}+C_{q^{\lambda}q_{\mu}}q^{\lambda}q_{\mu}+C_{g^{\lambda}_{\mu}}g^{\lambda}_{\mu}\nonumber\\
 &&+C_{q^{\lambda}\gamma_{\mu}\slashed q}q^{\lambda}\gamma_{\mu}\slashed q+C_{q^{\lambda}v_{\mu}\slashed q}q^{\lambda}v_{\mu}\slashed q+C_{q^{\lambda}q_{\mu}\slashed q}q^{\lambda}q_{\mu}\slashed q+C_{g^{\lambda}_{\mu}\slashed q}g^{\lambda}_{\mu}\slashed q\},
 \end{eqnarray}
where $C_{\rm structure}$ denotes the corresponding coefficient functions for different structure,
and are expressed as
\begin{eqnarray}
C_{q^{\lambda}\gamma_{\mu}}&=&\int_{0}^{2s_0} d \omega \int_{0}^{1} d u\Big\{
  -\sqrt{3} {\bar{u}}\big[{f^{(1)}} {\bar{u}} [2
({\hat{\psi}^{\bar{n}*}_{\parallel}}+{\hat{\psi}^{n*}_{\parallel}})+{\tilde{\psi}^{n*}_{\parallel}} \omega]-{f^{(2)}}
{m_{Q}} {\tilde{\psi}^{1*}_{\parallel}}\big]\frac{1}{\Delta^2}\nonumber\\
&&+2\sqrt{3}{\bar{u}}^2 \big[{f^{(1)}}
({\hat{\psi}^{\bar{n}*}_{\parallel}}+{\hat{\psi}^{n*}_{\parallel}}) \left(q^2+{\bar{u}} \omega  (2
{q\cdot v}+{\bar{u}} \omega )\right)-2 {f^{(2)}} {m_{Q}}
{\hat{\psi}^{n\bar{n}*}_{\parallel}} {q\cdot v}
\big]
{\frac{1}{\Delta^3}}\Big\}, \label{eq:cgamu} \\
C_{q^{\lambda}q_{\mu}}&=&\int_{0}^{2s_0} d \omega \int_{0}^{1} d u\Big\{ 2 \sqrt{3} {f^{(2)}} {\tilde{\psi}^{1*}_{\parallel}} {\bar{u}}{\frac{1}{\Delta}}
-4 \sqrt{3} {\bar{u}}^2 \Big[{f^{(1)}} {m_{Q}} ({\hat{\psi}^{n*}_{\parallel}}-{\hat{\psi}^{\bar{n}*}_{\parallel}})\nonumber\\
&&+{f^{(1)}} {\bar{u}} \omega  ({\hat{\psi}^{\bar{n}*}_{\parallel}}+{\hat{\psi}^{n*}_{\parallel}})-2 {f^{(2)}} {\hat{\psi}^{n\bar{n}*}_{\parallel}} ({m_{Q}}+{q\cdot v}+{\bar{u}} \omega )\Big]{\frac{1}{\Delta^2}}\Big\},\label{eq:cqmu2}
\\
C_{q^{\lambda}v_{\mu}}&=&\int_{0}^{2s_0} d \omega \int_{0}^{1} d u\Big\{
2 \sqrt{3} {\bar{u}}  \big[{f^{(1)}} {\tilde{\psi}^{n*}_{\parallel}} ({m_{Q}}+{\bar{u}}
\omega )+{f^{(2)}} {\bar{u}} (5{\hat{\psi}^{n\bar{n}*}_{\parallel}}+{\tilde{\psi}^{1*}_{\parallel}} \omega
)\big]
{\frac{1}{\Delta^2}}\nonumber\\
&&-4 \sqrt{3} {\bar{u}}^2 \left[{f^{(1)}} {\bar{u}} \omega  [{m_{Q}}
({\hat{\psi}^{n*}_{\parallel}}-{\hat{\psi}^{\bar{n}*}_{\parallel}})+{\bar{u}} \omega
({\hat{\psi}^{\bar{n}*}_{\parallel}}+{\hat{\psi}^{n*}_{\parallel}})]+2 {f^{(2)}} {\hat{\psi}^{n\bar{n}*}_{\parallel}}
\left(q^2+{q\cdot v} {\bar{u}} \omega \right)\right]
{\frac{1}{\Delta^3}}\Big\},\label{eq:cvmu}\nonumber\\
\\
C_{g^{\lambda}_{\mu}}&=&\int_{0}^{2s_0} d \omega \int_{0}^{1} d u\Big\{ -\sqrt{3} {\bar{u}} ({f^{(1)}} {\tilde{\psi}^{n*}_{\parallel}}+{f^{(2)}} {\tilde{\psi}^{1*}_{\parallel}}){\frac{1}{\Delta}}
+\Big[\frac{\sqrt{3}}{2} {\bar{u}}^2 [-{f^{(1)}} {m_{Q}} (2 {\hat{\psi}^{\bar{n}*}_{\parallel}}+{\hat{\psi}^{n*}_{\parallel}})\nonumber\\
&&+2 {f^{(1)}} {\bar{u}} \omega  ({\hat{\psi}^{\bar{n}*}_{\parallel}}+{\hat{\psi}^{n*}_{\parallel}})-4 {f^{(2)}} {\hat{\psi}^{n\bar{n}*}_{\parallel}} ({q\cdot v}+{\bar{u}} \omega)]
\Big]{\frac{1}{\Delta^2}}\Big\},\label{eq:cqmu1}
\\
C_{q^{\lambda}\gamma_{\mu}q\!\!\!\slash}&=&\int_{0}^{2s_0} d \omega \int_{0}^{1} d u\Big\{
\frac{\sqrt{3}}{2}  {f^{(1)}} {\tilde{\psi}^{n*}_{\parallel}} {\bar{u}}
{\frac{1}{\Delta^2}}
-4 \sqrt{3} {f^{(2)}} {m_{Q}} {\hat{\psi}^{n\bar{n}*}_{\parallel}} {\bar{u}}^2
{\frac{1}{\Delta^3}}\Big\},\label{eq:cgamuqslash}
\\
C_{q^{\lambda}q_{\mu}q\!\!\!\slash}&=&\int_{0}^{2s_0} d \omega \int_{0}^{1} d u\Big\{
-4 \sqrt{3} {f^{(1)}} {\bar{u}}^2 ({\hat{\psi}^{\bar{n}*}_{\parallel}}+{\hat{\psi}^{n*}_{\parallel}})
{\frac{1}{\Delta^3}}\Big\},\label{eq:cqmuqslash}
\\
C_{q^{\lambda}v_{\mu}q\!\!\!\slash}&=&\int_{0}^{2s_0} d \omega \int_{0}^{1} d u\Big\{
-4 \sqrt{3} {f^{(1)}} {\bar{u}}^3 \omega  ({\hat{\psi}^{\bar{n}*}_{\parallel}}+{\hat{\psi}^{n*}_{\parallel}}){\frac{1}{\Delta^3}}\Big\},\label{eq:cvmuqslash}
\\
C_{g^{\lambda}_{\mu}q\!\!\!\slash}&=&\int_{0}^{2s_0} d \omega \int_{0}^{1} d u\Big\{
 \sqrt{3} {f^{(1)}} {\bar{u}}^2 ({\hat{\psi}^{\bar{n}*}_{\parallel}}+{\hat{\psi}^{n*}_{\parallel}}){\frac{1}{\Delta^2}}\Big\}.\label{eq:cgmuqslash}
\end{eqnarray}
Here the arguments $(\omega,u)$ of the LCDAs $\psi_{\parallel}$, $\tilde{\psi}_{\parallel}$ and $\hat{\psi}_{\parallel}$ have been omitted to get terseness functions.
The correlation function at QCD level can be depicted by a Feynman diagram shown with Fig.~\ref{fig:FeymDiag} which is a function of $(p_{\Sigma^*}+q)^2$ and $q^2$.
By computing its discontinuity across the branching cut in the  complex plane of $(p_{\Sigma^*}+q)^2$, we can transform the correlation function Eq.(\ref{correQCD})  into a dispersion integration form,
\begin{eqnarray}
\Pi_{\mu,V}^{QCD}(p_{\Sigma^*},q)=\frac{1}{\pi}\int_{(m_{Q}+m_{Q^{\prime}}+m_{q})^{2}}^{sth}ds\frac{\rm{Im}\Pi_{\mu,V}^{QCD}(s,q^2)}{s-(p_{\Sigma^*}+q)^{2}}.\label{DiscorreQCD}
\end{eqnarray}
Based on the global Quark-Hadron duality, there would exist an equivalence between the QCD contribution Eq.~(\ref{DiscorreQCD}) in spectral region $(m_{Q}+m_{Q^{\prime}}+m_{q})^{2}<s<sth$ and the continuum contribution Eq.~(\ref{correHadron}). Therefore, Eq.~(\ref{DiscorreQCD}) is equal to Eq.~(\ref{FormextractP}), and then the form factors $F_i^{+}$ can be extracted out as
\begin{eqnarray}
& &-\frac{f_{\Xi}\bar{u}_{\alpha}(p_{\Sigma^*})F_{1}^{+}(q^{2})}{(q+p_{\Sigma^*})^{2}-m_{\Xi}^{2}} =  \frac{1}{\pi}\int_{(m_{Q}+m_{Q^{\prime}}+m_{q})^{2}}^{sth}ds\frac{{\rm Im}[C_{q^{\lambda}\gamma_{\mu}}+C_{q^{\lambda}\gamma_{\mu}\slashed q} (m_{\Sigma^*}-m_{\Xi})]}{2 m_{\Xi}[s-(p_{\Sigma^*}+q)^{2}]},\label{fromfactor1} \nonumber\\
& &-\frac{f_{\Xi}\bar{u}_{\alpha}(p_{\Sigma^*})F_{2}^{+}(q^{2})}{(q+p_{\Sigma^*})^{2}-m_{\Xi}^{2}}= \frac{1}{\pi}\int_{(m_{Q}+m_{Q^{\prime}}+m_{q})^{2}}^{sth}ds\frac{{\rm Im}[-2 m_{\Sigma^*}C_{q^{\lambda}\gamma_{\mu}\slashed q} +(m_{\Sigma^*}^2+m_{\Sigma^*}m_{\Xi}) C_{q^{\lambda}q_{\mu}\slashed q}}{2 m_{\Sigma^*}m_{\Xi}[s-(p_{\Sigma^*}+q)^{2}]}\nonumber\\
&&\qquad\qquad\qquad\qquad\qquad\qquad\qquad\qquad\qquad-(m_{\Xi}+m_{\Sigma^*})C_{q^{\lambda}v_{\mu}\slashed q}-m_{\Sigma^*} C_{q^{\lambda}q_{\mu}}+C_{q^{\lambda}v_{\mu}}],\label{fromfactor2}\nonumber\\
& &-\frac{f_{\Xi}\bar{u}_{\alpha}(p_{\Sigma^*})F_{3}^{+}(q^{2}) }{(q+p_{\Sigma^*})^{2}-m_{\Xi}^{2}}=  \frac{1}{\pi}\int_{(m_{Q}+m_{Q^{\prime}}+m_{q})^{2}}^{sth}ds\frac{{\rm Im}[C_{q^{\lambda}q_{\mu}}-(m_{\Xi}+m_{\Sigma^*}) C_{q^{\lambda}q_{\mu}\slashed q}]}{2 m_{\Xi}[s-(p_{\Sigma^*}+q)^{2}]},\label{fromfactor3} \nonumber\\
& &-\frac{f_{\Xi}\bar{u}_{\alpha}(p_{\Sigma^*})F_{4}^{+}(q^{2})}{(q+p_{\Sigma^*})^{2}-m_{\Xi}^{2}} =  \frac{1}{\pi}\int_{(m_{Q}+m_{Q^{\prime}}+m_{q})^{2}}^{sth}ds\frac{{\rm Im}[C_{g^{\lambda}_{\mu}}-(m_{\Xi}+m_{\Sigma^*}) C_{g^{\lambda}_{\mu}\slashed q}]}{2 m_{\Xi}[s-(p_{\Sigma^*}+q)^{2}]},\label{fromfactor4}
\end{eqnarray}
which correspond to the transition of $\Xi^{P+}_{QQ^{\prime}}$. In order to suppress the contribution from higher twists, the Borel transformation is performed on both sides of three Eqs.~(\ref{fromfactor1})-(\ref{fromfactor4}). To compute the discontinuity, the following transformation on the high power denominators in Eqs.~(\ref{eq:cgamu})-(\ref{eq:cvmuqslash}) has to be conducted before doing the Borel transformation,
\begin{eqnarray}
{\frac{1}{[(q+\bar{u}\omega v)^2-m_{Q}^2]^n}}\to\frac{1}{(n-1)!}\left(\frac{\partial}{\partial \Omega}\right)^{(n-1)}{\frac{1}{(q+\bar{u}\omega v)^2-\Omega}}\Big|_{\Omega=m_Q^2}.
\end{eqnarray}
After that, the form factors $F_i^{+}$ can be written as a series of explicit expressions.
Here we just show $F_{1}^{+}$ as an example,
\begin{eqnarray}
& &F_{1}^{+}(q^{2}) =
-\frac{1}{f_{\Xi}}exp(\frac{m_{\Xi}^2}{M^2}) \frac{\partial}{\partial \Omega}
\int_{0}^{2s_0} d \omega \int_{0}^{1} d u\frac{m_{\Sigma^*}}{\bar{u}\omega}exp(-\frac{s_{r}^{\Omega}}{M^2})
\theta(sth-s_{r}^{\Omega})\theta(s_{r}^{\Omega}-(\sqrt{\Omega}+m_{Q^{\prime}}+m_{q})^{2})\nonumber\\
&&\times\Big\{-\sqrt{3} {\bar{u}}\big[{f^{(1)}} {\bar{u}} [2
({\hat{\psi}^{\bar{n}*}_{\parallel}}+{\hat{\psi}^{n*}_{\parallel}})+{\tilde{\psi}^{n*}_{\parallel}} \omega]-{f^{(2)}}
{m_{Q}} {\tilde{\psi}^{1*}_{\parallel}}\big]\frac{1}{2m_{\Xi}}
+\frac{\sqrt{3}}{2}  {f^{(1)}} {\tilde{\psi}^{n*}_{\parallel}} {\bar{u}}\frac{m_{\Sigma^*}-m_{\Xi}}{2m_{\Xi}}\Big\}\Big|_{\Omega=m_Q^2}\nonumber\\
  &&-\frac{1}{f_{\Xi}}exp(\frac{m_{\Xi}^2}{M^2}) \frac{1}{2}\left(\frac{\partial}{\partial \Omega}\right)^2
\int_{0}^{2s_0} d \omega \int_{0}^{1} d u\frac{m_{\Sigma^*}}{\bar{u}\omega}exp(-\frac{s_{r}^{\Omega}}{M^2})
\theta(sth-s_{r}^{\Omega})\theta(s_{r}^{\Omega}-(m_{Q}+m_{Q^{\prime}}+m_{q})^{2})\nonumber\\
&&\times\Big\{2\sqrt{3}{\bar{u}}^2 \big[{f^{(1)}}
({\hat{\psi}^{\bar{n}*}_{\parallel}}+{\hat{\psi}^{n*}_{\parallel}}) \left(q^2+{\bar{u}} \omega  (2
{q\cdot v}+{\bar{u}} \omega )\right)-2 {f^{(2)}} {m_{Q}}
{\hat{\psi}^{n\bar{n}*}_{\parallel}} {q\cdot v}
\big]\frac{1}{2m_{\Xi}}\nonumber\\
&&\quad-4 \sqrt{3} {f^{(2)}} {m_{Q}} {\hat{\psi}^{n\bar{n}*}_{\parallel}} {\bar{u}}^2 \frac{m_{\Sigma^*}-m_{\Xi}}{2m_{\Xi}}\Big\}\Big|_{\Omega=m_Q^2},   \label{fromfactor11}
  \end{eqnarray}
where $s_r$ and $s^{\Omega}_r$ are defined as the singularity position of the heavy quark propagator in the correlation function. In other words, they are the roots of the following two equations respectively
\begin{eqnarray}
&&\frac{\bar{u}\omega}{m_{\Sigma^*}}s_r+H(u,\omega,q^{2})-m_{Q}^2=0,\nonumber\\
&&\frac{\bar{u}\omega}{m_{\Sigma^*}}s^{\Omega}_r+H(u,\omega,q^{2})-\Omega=0.\label{eq:srandq}
\end{eqnarray}
With similar procedure one can get expressions of $G_i^{+}$, which will not be repeated here.

\section{Numerical results}
\label{sec:numerical results}
\subsection{Transition Form Factors}

In the numerical analysis, we take the heavy quark masses as $m_{c}=(1.35\pm0.10)\ {\rm GeV}$ and $m_{b}=(4.7\pm0.1)\ {\rm GeV}$, and neglect the masses of light quarks. The input parameters (mass lifetime, decay constant) of doubly heavy baryons are listed in Tab.~\ref{Tab:para_if}~\cite{Karliner:2014gca,Shah:2016vmd,Shah:2017liu,Kiselev:2001fw}. Moreover, the masses $\Sigma_Q^*$ are taken as $m_{\Sigma_c^*}=2.454$ GeV, $m_{\Sigma_b^*}=5.814$ GeV, and decay constants of them are $f^{(1)}=f^{(2)}=0.038$~\cite{Groote:1997yr}. The upper limit of light quarks momentum is $s_0=1.2$ GeV~\cite{Ali:2012pn}.

\begin{table}[!htb]
\caption{The input parameters of doubly heavy baryons and the fitting parameters for transition form factors.}
\label{Tab:para_if} \footnotesize%
\begin{tabular}{c|c|c|c|c|c|c|c}
\hline
\hline
Baryons & Mass (GeV) & Lifetime (fs) &$f_{\Xi}$ $({\rm GeV}^{3})$~\cite{Hu:2017dzi}&Channel& $s_{\rm th}$ (GeV$^{2}$) & $M^{2}$ (GeV$^{2}$) & Fit Range (GeV$^{2}$)\tabularnewline
\hline
$\Xi_{cc}^{++}$  &\multirow{2}{*}{ $3.621$ \cite{Aaij:2017ueg}}  & 256 ~\cite{Aaij:2018wzf}  &\multirow{2}{*}{$0.109\pm0.021$}&\multirow{2}{*}{$\Xi_{cc}\to\Sigma^{*}_{c}$}&\multirow{2}{*}{$16\pm\textcolor{red}{2}$}&\multirow{2}{*}{$15\pm1$}&\multirow{2}{*}{$0<q^{2}<0.8$} \tabularnewline
$\Xi_{cc}^{+}$  &  & 45~\cite{Cheng:2018mwu} &  &&&&\tabularnewline\hline
$\Xi_{bc}^{+}$  &\multirow{2}{*}{ $6.943$ \cite{Brown:2014ena}}  & 244  \cite{Karliner:2014gca}&\multirow{2}{*}{ $0.176\pm0.040$} &$\Xi_{bc}\to\Sigma^{*}_{b}$ & $54\pm\textcolor{red}{3}$ & $20\pm1$ & $0<q^{2}<0.8$\tabularnewline
$\Xi_{bc}^{0}$  &  & 93 \cite{Karliner:2014gca}&  &$\Xi_{bc}\to\Sigma^{*}_{c}$& $54\pm\textcolor{red}{3}$ & $20\pm1$ & $0<q^{2}<6$\tabularnewline\hline
$\Xi_{bb}^{0}$  & \multirow{2}{*}{$10.143$ \cite{Brown:2014ena}}  & 370 \cite{Karliner:2014gca}& \multirow{2}{*}{$0.281\pm0.071$} &\multirow{2}{*}{$\Xi_{bb}\to\Sigma^{*}_{b}$}&\multirow{2}{*}{$112\pm\textcolor{red}{5}$}&\multirow{2}{*}{$20\pm1$}&\multirow{2}{*}{ $0<q^{2}<6$}\tabularnewline
$\Xi_{bb}^{-}$  &   & 370 \cite{Karliner:2014gca}&  &&&&\tabularnewline
\hline
\hline
\end{tabular}
\end{table}
The Borel parameters $M^2$ and threshold $s_{\rm th}$ of $\Xi_{QQ^{\prime}}$ are  carefully chosen as shown in Tab.~\ref{Tab:para_if}. Here we only choose the uncertainty of Borel parameter $M$ and thresholds $s_{\rm {th}}$ to enumerate the error of resulting form factors.  We take the uncertainty of Borel parameters around their center value as $1\ {\rm GeV}^2$ while the uncertainty of $s_{\rm {th}}$ as $2\sim 5\ {\rm GeV}^2$.
Using these input parameters, we can obtain the numerical result of the transition form factors $f_{i}$/$g_{i}$($i=1,2,3,4$) defined by Eq.(\ref{eq:parameterization1}). 

It should be mentioned that in Ref.~\cite{Ali:2012pn}, the LCDAs depend on an parameter $A$ ranging from 0 to 1. In principle $A$ should be taken as its center value $0.5$.  However, with the use of the LCDA model given in Ref.~\cite{Ali:2012pn}, some of the LCDAs will be divergent if $A=0.5$. To make the LCDAs be finite one has to choose a smaller value of $A$. In this work, to ensure the stability of the  from factors,  the value of $A$ closest to its center value is chosen as $0.21$.   

Since the LCSR approach used here is only reliable in the small $q^2$ region, one has to extend the form factors in this region to the whole physical region $0\sim (M_{\Xi_{QQ^{\prime}}}-M_{\Sigma_{Q^{\prime}}})^2$. To realize this extension we can choose a parameterization function to fit the form factors in the small $q^2$ region, and then the obtained parameterization  function can describe the form factors in the whole physical region. The small $q^2$ regions chosen for the fitting are listed in the last column of Tab.~\ref{Tab:para_if}.
The parameterization function is chosen as the dipole form
\begin{equation}
	F(q^{2})=\frac{F(0)[1+a (q^2)+b (q^2)^2]}{1-\frac{q^{2}}{m_{{\rm fit}}^{2}}+\delta\left(\frac{q^{2}}{m_{{\rm fit}}^{2}}\right)^{2}}\label{eq:fit_formula_1}.
\end{equation}
The numerical results of the transition form factors are listed in Tab.~\ref{Tab:ff_all}. There are two further points should be mentioned:
\begin{itemize} 
    \item  In the Tab.~\ref{Tab:ff_all}, the sign ``$-$" refers to the corresponding parameter without fitting, and the asterisk at the upper right corner of the data indicates the use of  fitting formula: 
    \begin{equation}
      F(q^{2})=\frac{F(0)[1+a (q^2)+b (q^2)^2]}{1+\frac{q^{2}}{m_{{\rm fit}}^{2}}+\delta\left(\frac{q^{2}}{m_{{\rm fit}}^{2}}\right)^{2}}\label{eq:fit_formula_2}.
    \end{equation}    
    \item  As shown in the Tab.~\ref{Tab:ff_all},  the fitting parameters of some form factors are abnormally large. Although these large values are unexpected, with these parameters the behavior the form factors are still stable in the physical range. 
\end{itemize}

\begin{figure}
\includegraphics[width=0.8\columnwidth]{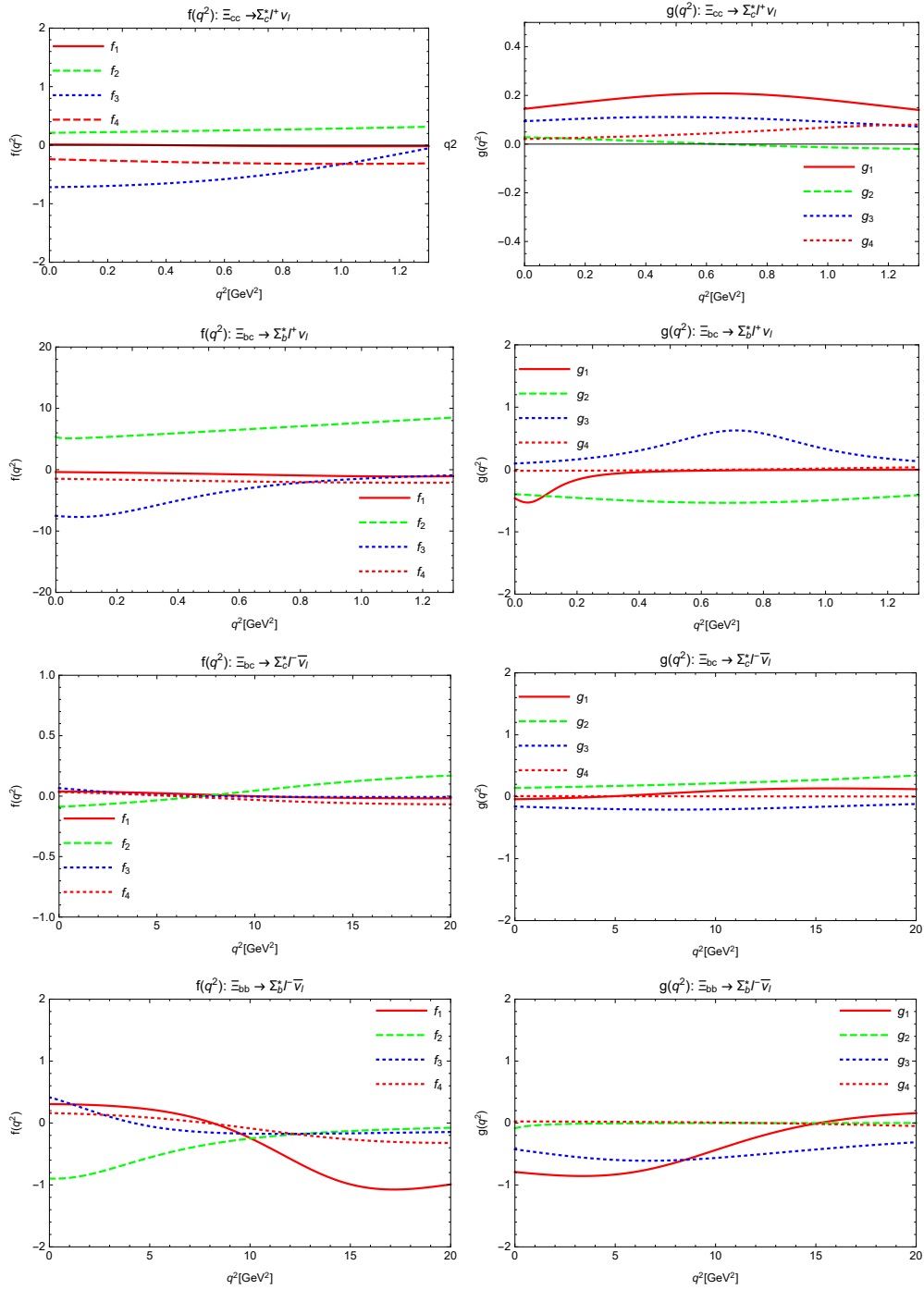}
\caption{$q^2$ dependence of the $\Xi_{QQ^{\prime}q_{2}}\to \Sigma^{*}_{Q^{\prime}q_{1}q_{2}}$ form factors. The first two graphs correspond to $\Xi_{cc}\to\Sigma^{*}_c$, the second two graphs correspond to $\Xi_{bc}\to\Sigma^{*}_b$, the third two graphs correspond to $\Xi_{bc}\to\Sigma^{*}_c$ and the fourth two graphs correspond to $\Xi_{bb}\to\Sigma^{*}_b$. Here the parameters $s_{\rm th}$, $M^2$ are fixed at their center values as shown in Tab.~\ref{Tab:para_if}. For the case of $\Xi_{QQ^{\prime}q}\to\Sigma^{*}_{Q^{\prime}qq}$, the vertical scale needs to be enlarged by a factor $\sqrt{2}$.}
\label{fig:Formfactors}
\end{figure}

\begin{table}
  \caption{Numerical results of transition form factors of the $\Xi_{QQ^{\prime}}\to\Sigma^{*}_{Q^{\prime}}$.
$F(0)$, $m_{\rm{fix}}$, $\delta$, $a$ and $b$ correspond to the five
fitting parameters in Eq.~(\ref{eq:fit_formula_1}). The form factors
of $\Xi_{QQ^{\prime}q}\to\Sigma^{*}_{Q^{\prime}qq}$ are just $\sqrt{2}$
times those of $\Xi_{QQ^{\prime}q_{2}}\to\Sigma^{*}_{Q^{\prime}q_{1}q_{2}}$,
which are not listed in this table.}
\label{Tab:ff_all}
\begin{tabular}{c|c|c|c|c|c}
  \hline\hline
  $F$ &$F(0)$ &$m_{\rm{fit}}$ &$\delta$ &$a$ &$b$\\ \hline
  $f_{1}^{\Xi_{cc}\to\Sigma_{c}^{*}}$ &$
  0.01\pm0.01$ &$-0.77\pm1.64$ &$0.56\pm0.41$ &$-2.51\pm3.07$ &$-$ \\
  $f_{2}^{\Xi_{cc}\to\Sigma_{c}^{*}}$ &$
  0.21\pm0.04$ &$1.88\pm0.07$ &$0.28\pm0.49$ &$-$ &$-$ \\
  $f_{3}^{\Xi_{cc}\to\Sigma_{c}^{*}}$ &$
  -0.72\pm0.09$ &$-1.23\pm2.57$ &$0.53\pm0.49$ &$-0.74\pm0.10$ &$-$ \\
  $f_{4}^{\Xi_{cc}\to\Sigma_{c}^{*}}$ &$
  -0.24\pm0.04$ &$1.41\pm2.88$ &$1.00\pm0.11$ &$-$ &$-$ \\
  $g_{1}^{\Xi_{cc}\to\Sigma_{c}^{*}}$ &$
  0.14\pm0.02$ &$1.02\pm2.08$ &$0.82\pm0.09$ &$-$ &$-$ \\
  $g_{2}^{\Xi_{cc}\to\Sigma_{c}^{*}}$ &$
  0.03\pm0.01$ &$-1.52\pm4.01$ &$3.27\pm2.05$ &$-1.55\pm0.27$ &$-$ \\
  $g_{3}^{\Xi_{cc}\to\Sigma_{c}^{*}}$ &$
  0.09\pm0.03$ &$-1.24\pm2.76$ &$1.59\pm2.65$ &$-$ &$-$ \\
  $g_{4}^{\Xi_{cc}\to\Sigma_{c}^{*}}$ &$
  0.02\pm0.01$ &$-0.95\pm1.98$ &$0.34\pm0.01$ &$-$ &$-$ \\ 
     \hline
  $f_{1}^{\Xi_{bc}\to\Sigma_{b}^{*}}$ &$
  -0.38\pm0.09$ &$-0.94\pm1.83$ &$0.38\pm0.01$ &$-$ &$-$ \\
  $f_{2}^{\Xi_{bc}\to\Sigma_{b}^{*}}$ &$
  5.36\pm0.75^{*}$ &$0.15\pm0.05^{*}$ &$0.30\pm0.15^{*}$ &$38.30\pm30.70^{*}$ &$
    22.40\pm10.80^{*}$ \\
  $f_{3}^{\Xi_{bc}\to\Sigma_{b}^{*}}$ &$
  -7.51\pm0.72$ &$-1.20\pm0.15$ &$9.93\pm5.74$ &$-$ &$-$ \\
  $f_{4}^{\Xi_{bc}\to\Sigma_{b}^{*}}$ &$
  -1.47\pm0.15$ &$1.41\pm0.03$ &$0.81\pm0.01$ &$-$ &$-$ \\
  $g_{1}^{\Xi_{bc}\to\Sigma_{b}^{*}}$ &$
  -0.46\pm0.08$ &$0.40\pm0.92$ &$1.94\pm1.07$ &$-$ &$-$ \\
  $g_{2}^{\Xi_{bc}\to\Sigma_{b}^{*}}$ &$
  -0.40\pm0.04$ &$-1.14\pm2.33$ &$0.96\pm0.19$ &$-$ &$-$ \\
  $g_{3}^{\Xi_{bc}\to\Sigma_{b}^{*}}$ &$
  0.10\pm0.06$ &$-0.65\pm0.04$ &$0.30\pm0.03$ &$-$ &$-$ \\
  $g_{4}^{\Xi_{bc}\to\Sigma_{b}^{*}}$ &$
  -0.02\pm0.01$ &$1.05\pm2.20$ &$0.45\pm0.22$ &$-1.34\pm0.23$ &$-$ \\
   \hline
  $f_{1}^{\Xi_{bc}\to\Sigma_{c}^{*}}$ &$
  0.04\pm0.00$ &$3.47\pm6.62$ &$1.03\pm15.30$ &$-0.10\pm0.07$ &$-$ \\
  $f_{2}^{\Xi_{bc}\to\Sigma_{c}^{*}}$ &$
  -0.09\pm0.01$ &$4.46\pm1.10$ &$0.90\pm2.50$ &$-0.14\pm0.02$ &$-$ \\
  $f_{3}^{\Xi_{bc}\to\Sigma_{c}^{*}}$ &$
  0.07\pm0.01^{*}$ &$4.54\pm3.55^{*}$ &$15.50\pm18.50^{*}$ &$-0.13\pm0.16^{*}$ &$-$ \\
  $f_{4}^{\Xi_{bc}\to\Sigma_{c}^{*}}$ &$
  0.03\pm0.00$ &$4.17\pm8.47$ &$1.02\pm0.92$ &$-0.17\pm0.02$ &$-$ \\
  $g_{1}^{\Xi_{bc}\to\Sigma_{c}^{*}}$ &$
  -0.04\pm0.01$ &$-3.19\pm6.41$ &$0.55\pm0.29$ &$-0.23\pm0.05$ &$-$ \\
  $g_{2}^{\Xi_{bc}\to\Sigma_{c}^{*}}$ &$
  0.14\pm0.01$ &$5.09\pm0.02$ &$0.31\pm0.03$ &$-$ &$-$ \\
  $g_{3}^{\Xi_{bc}\to\Sigma_{c}^{*}}$ &$
  -0.16\pm0.02$ &$4.02\pm0.15$ &$1.03\pm0.02$ &$-$ &$-$ \\
  $g_{4}^{\Xi_{bc}\to\Sigma_{c}^{*}}$ &$
  0.01\pm0.00^{*}$ &$5.06\pm0.16^{*}$ &$1.19\pm0.50^{*}$ &$-$ &$-$ \\ \hline
  $f_{1}^{\Xi_{bb}\to\Sigma_{b}^{*}}$ &$
  0.31\pm0.03$ &$2.95\pm0.12$ &$0.33\pm0.01$ &$-0.12\pm0.00$ &$-$ \\
  $f_{2}^{\Xi_{bb}\to\Sigma_{b}^{*}}$ &$
  -0.90\pm0.10$ &$8.45\pm1.21$ &$139.00\pm93.30$ &$-$ &$-$\\
  $f_{3}^{\Xi_{bb}\to\Sigma_{b}^{*}}$ &$
  0.42\pm0.05$ &$6.89\pm0.60$ &$57.60\pm58.10$ &$-0.24\pm0.02$ &$-$ \\
  $f_{4}^{\Xi_{bb}\to\Sigma_{b}^{*}}$ &$
  0.16\pm0.01$ &$-3.42\pm7.07$ &$0.51\pm0.16$ &$-0.13\pm0.01$ &$-$ \\
  $g_{1}^{\Xi_{bb}\to\Sigma_{b}^{*}}$ &$
  -0.79\pm0.11$ &$3.06\pm0.98$ &$0.60\pm1.31$ &$-0.07\pm0.05$ &$-$ \\
  $g_{2}^{\Xi_{bb}\to\Sigma_{b}^{*}}$ &$
  -0.09\pm0.01^{*}$ &$1.04\pm0.07^{*}$ &$0.36\pm0.62^{*}$ &$-$ &$-$ \\
  $g_{3}^{\Xi_{bb}\to\Sigma_{b}^{*}}$ &$
  -0.42\pm0.06$ &$4.01\pm0.08$ &$1.78\pm0.10$ &$0.04\pm0.00$ &$-$ \\
  $g_{4}^{\Xi_{bb}\to\Sigma_{b}^{*}}$ &$
  0.03\pm0.00$ &$-4.21\pm0.08$ &$0.37\pm0.09$ &$-0.08\pm0.00$ &$-$ \\ 
     \hline
  \hline
  \end{tabular}
  \end{table}

\subsection{Semi-leptonic Decays}
For the semileptonic decays $\Xi_{QQ^{\prime}}\to\Sigma^{*}_{Q^{\prime}}l\nu_{l}$ considered in this work, the effective Hamiltonian is the same with the ones of $\Xi_{QQ^{\prime}}\to\Sigma_{Q^{\prime}}l\nu_{l}$ in our previous work~\cite{Hu:2019bqj},
\begin{eqnarray}
{\cal H}_{{\rm eff}} & = & \frac{G_{F}}{\sqrt{2}}\big\{V_{ub}[\bar{u}\gamma_{\mu}(1-\gamma_{5})b][\bar{l}\gamma^{\mu}(1-\gamma_{5})\nu_{l}]+V_{cd}^{*}[\bar{d}\gamma_{\mu}(1-\gamma_{5})c][\bar{\nu}_{l}\gamma^{\mu}(1-\gamma_{5})l]\big\}.
\end{eqnarray}
Here the Fermi constant $G_{F}=1.166\times10^{-5}{\rm GeV}^{-2}$, and CKM matrix elements $|V_{ub}|=0.00357$, $|V_{cd}|=0.225$~\cite{Olive:2016xmw,Tanabashi:2018oca}.

After a thorough analysis on the polarizations of the initial and final states, the positive helicity amplitudes induced by V-A current can be derived as
\begin{eqnarray}
&&H_{{3/2},1}^{V} =  -i\sqrt{Q_{-}}f_{4},\;
H_{\frac{1}{2},1}^{V} =  i\sqrt{\frac{Q_{-}}{3}}\Big(f_{4}-\frac{Q_{+}}{M_{1}M_{2}}f_{1}\Big),\; H_{-1/2,t}^{V}=i\sqrt{\frac{2}{3}}\frac{\sqrt{Q_{+}}Q_{-}(M_{1}^{2}-M_{2}^{2})}{2M_{1}^{2}M_{2}}f_{3},\nonumber \\
&&H_{{3/2},1}^{A} = i\sqrt{Q_{+}}g_{4},\;
H_{\frac{1}{2},1}^{A} =  i\sqrt{\frac{Q_{+}}{3}}\Big(g_{4}-\frac{Q_{-}}{M_{1}M_{2}}g_{1}\Big),\; H_{-1/2,t}^{A}=-i\sqrt{\frac{2}{3}}\frac{\sqrt{Q_{-}}Q_{+}(M_{1}^{2}-M_{2}^{2})}{2M_{1}^{2}M_{2}}g_{3},\nonumber \\
&&H_{{1/2},0}^{V}  =  i\sqrt{\frac{2}{3}}\frac{\sqrt{Q_{-}}}{\sqrt{q^{2}}}\Big[\frac{M_{1}^{2}-M_{2}^{2}-q^2}{2M_{2}}f_{4}-\frac{M_{1}-M_{2}}{2M_{1}M_{2}}Q_{+}f_{1}-\frac{Q_{+}Q_{-}}{2M_{1}^{2}M_{2}}f_{2}\Big],\nonumber \\
&&H_{{1/2},0}^{A}  =  -i\sqrt{\frac{2}{3}}\frac{\sqrt{Q_{+}}}{\sqrt{q^{2}}}\Big[\frac{M_{1}^{2}-M_{2}^{2}-q^2}{2M_{2}}g_{4}+\frac{M_{1}-M_{2}}{2M_{1}M_{2}}Q_{-}g_{1}-\frac{Q_{+}Q_{-}}{2M_{1}^{2}M_{2}}g_{2}\Big].\label{eq:helicityampli}
\end{eqnarray}

The negative helicity amplitudes can be also derived, which have the following relations with the corresponding positive ones,
\begin{equation}
H_{-\lambda_{2},-\lambda_{W}}^{V}=-H_{\lambda_{2},\lambda_{W}}^{V}\quad\text{and}\quad H_{-\lambda_{2},-\lambda_{W}}^{A}=H_{\lambda_{2},\lambda_{W}}^{A},
\end{equation}
$\lambda_{2}$ and $\lambda_{W}$ denotes the polarizations of the final $\Sigma_{Q^{\prime}}$ and the intermediate $W$ boson, respectively. The total helicity amplitudes induced by the $V-A$ current are
\begin{equation}
H_{\lambda_{2},\lambda_{W}}=H_{\lambda_{2},\lambda_{W}}^{V}-H_{\lambda_{2},\lambda_{W}}^{A}.
\end{equation}


Considering longitudinally and transversely polarized $l\nu$ pairs, we can write the differential decay widths of $\Xi_{QQ^{\prime}}\to\Sigma^{*}_{Q^{\prime}}l\nu$ in two parts
\begin{align}
\frac{d\Gamma_{L}}{dq^{2}} & =\frac{G_{F}^{2}|V_{{\rm CKM}}|^{2}q^{2}\ p\ (1-\hat{m}_{l}^{2})^{2}}{384\pi^{3}M_{1}^{2}}\big[(2+\hat{m}_{l}^{2})(|H_{-\frac{1}{2},0}|^{2}+|H_{\frac{1}{2},0}|^{2})+3\hat{m}_{l}^{2}(|H_{-\frac{1}{2},t}|^{2}+|H_{\frac{1}{2},t}|^{2})\big],\label{eq:longi-1}\\
\frac{d\Gamma_{T}}{dq^{2}} & =\frac{G_{F}^{2}|V_{{\rm CKM}}|^{2}q^{2}\ p\ (1-\hat{m}_{l}^{2})^{2}(2+\hat{m}_{l}^{2})}{384\pi^{3}M_{1}^{2}}(|H_{\frac{1}{2},1}|^{2}+|H_{-\frac{1}{2},-1}|^{2}+|H_{\frac{3}{2},1}|^{2}+|H_{-\frac{3}{2},-1}|^{2}),\label{eq:trans-1}
\end{align}

In the above Eqs.~(\ref{eq:helicityampli})-(\ref{eq:trans-1}), $Q_{\pm}=(M_1\pm M_2)^{2}-q^{2}$, $M_{1}$ and $M_{2}$ refer to the masses of $\Xi_{QQ^{\prime}}$ and $\Sigma_{Q}$ respectively. $\hat{m}_{l}\equiv m_{l}/\sqrt{q^{2}}$, where $l=e,\mu,\tau$. In this work, the lepton masses $m_{e}$ and $m_{\mu}$ are neglected while $m_{\tau}=1.78$~GeV is taken from Ref.~\cite{Olive:2016xmw}.
Before integrating out the squared transfer momentum $q^{2}$, it is worth for us to note that the calculation of the decay width is conducted in the rest frame of the initial particle $\Xi_{QQ^{\prime}}$ for plainness. $p=\sqrt{Q_{+}Q_{-}}/(2M_{1})$ is the magnitude of three-momentum  of $\Sigma_{Q^{\prime}}^{*}$ in the rest frame of $\Xi_{QQ^{\prime}}$.
After the integration of $q^{2}$, the total decay width can be obtained
\begin{equation}
\Gamma=\int_{m_l^2}^{(M_{1}-M_{2})^{2}}dq^{2}\left(\frac{d\Gamma_{L}}{dq^{2}}+\frac{d\Gamma_{T}}{dq^{2}}\right).
\end{equation}
Our predictions for various semi-leptonic $\Xi_{QQ^{\prime}}\to\Sigma^{*}_{Q^{\prime}}l\nu_l$ processes, including their integrated partial decay widths, branching ratios and the ratios of $\Gamma_{L}/\Gamma_{T}$ are given in Tab.~\ref{Tab:semi_lep}.

Here are some comments on these phenomenology results in Tab.~\ref{Tab:semi_lep}:
\begin{itemize}
\item From Tab.~\ref{Tab:semi_lep}, the relatively large branching ratios of $\Xi_{bc}\to\Sigma_{b}^{*}l\nu$ and $\Xi_{bb}\to\Sigma_{b}^{*}l\bar{\nu}$ imply the potential for observing $\Xi_{bc},\  \Xi_{bb}$ through these decay channels in the future experiments.
\item The uncertainty of the decay widths are caused by the error of form factors which comes from the fitting parameters, the Borel parameter $M$ and the thresholds $s_{\rm th}$. 
\item Using the SU(3) flavor symmetry, we can obtain following model-independent relations among different decays, which leads to the semi-leptonic decay widths of $\Omega_{QQ^{\prime}}$~\cite{Wang:2017azm,Shi:2017dto,Hu:2020mxk}:
\begin{eqnarray}
\Gamma(\Omega_{cc}^{+}\to\Xi_{c}^{\prime*0}l^{+}\nu)&=&\Gamma(\Xi_{cc}^{++}\to\Sigma_{c}^{*+}l^{+}\nu)=\frac{1}{2}\Gamma(\Xi_{cc}^{+}\to\Sigma_{c}^{*0}l^{+}\nu)=
(  7.99\pm1.97)\times10^{-17} \rm{GeV},\nonumber\\
\Gamma(\Omega_{bc}^{0}\to\Xi_{b}^{\prime*-}l^{+}\nu)&=&\Gamma(\Xi_{bc}^{+}\to\Sigma_{b}^{*0}l^{+}\nu)=\frac{1}{2}\Gamma(\Xi_{bc}^{0}\to\Sigma_{b}^{*-}l^{+}\nu)=
(  1.93\pm0.44)\times10^{-17} \rm{GeV},\nonumber\\
\Gamma(\Omega_{bc}^{0}\to\Xi_{c}^{\prime*+}l^{-}\bar{\nu})&=&\Gamma(\Xi_{bc}^{0}\to\Sigma_{c}^{*+}l^{-}\bar{\nu})=\frac{1}{2}\Gamma(\Xi_{bc}^{+}\to\Sigma_{c}^{*++}l^{-}\bar{\nu})=
(2.87\pm0.52)\times10^{-17} \rm{GeV},\nonumber\\
\Gamma(\Omega_{bb}^{-}\to\Xi_{b}^{\prime*0}l^{-}\bar{\nu})&=&\Gamma(\Xi_{bb}^{-}\to\Sigma_{b}^{*0}l^{-}\bar{\nu})=\frac{1}{2}\Gamma(\Xi_{bb}^{0}\to\Sigma_{b}^{*+}l^{-}\bar{\nu})=
(7.72\pm1.80)\times10^{-17}\rm{GeV}.\nonumber\\\label{su3symmetry}
\end{eqnarray}
\end{itemize}

\begin{table}
  \caption{Decay widths and branching ratios of the semi-leptonic
  $\Xi_{QQ^{\prime}}\to\Sigma^{*}_{Q^{\prime}}l\nu_{l}$
   decays, where $l=e,\mu$}\label{Tab:semi_lep} %
   \begin{tabular}{l|c|c|c}
    \hline\hline
    channels &$\Gamma/\text{~GeV}$ &${\cal B}$ &$\Gamma_{T}/\Gamma_{L}$\\ \hline
    $\Xi_{cc}^{++}\to\Sigma_{c}^{*+}l^{+}\nu_{l} $ &$(
    7.99\pm1.97)\times 10^{-17}$ &
    $(
    3.11\pm0.77)\times 10^{-5}$ &
    $0.293\pm0.165$\\
    $\Xi_{cc}^{+}\to\Sigma_{c}^{*0}l^{+}\nu_{l} $ &$(
    1.59\pm0.39)\times 10^{-16}$ &
    $(
    1.09\pm0.27)\times 10^{-5}$ &
    $0.293\pm0.165$\\
    $\Xi_{bc}^{+}\to\Sigma_{b}^{*0}l^{+}\nu_{l} $ &$(
    1.93\pm0.44)\times 10^{-16}$ &
    $(
    7.14\pm1.64)\times 10^{-5}$ &
    $0.339\pm0.125$\\
    $\Xi_{bc}^{0}\to\Sigma_{b}^{*-}l^{+}\nu_{l} $ &$(
    3.79\pm0.87)\times 10^{-16}$ &
    $(
    5.36\pm1.23)\times 10^{-5}$ &
    $0.339\pm0.125$\\
    $\Xi_{bc}^{0}\to\Sigma_{c}^{*+}l^{-}\bar{\nu}_{l} $ &$(
    1.47\pm0.32)\times 10^{-17}$ &
    $(
    2.08\pm0.46)\times 10^{-6}$ &
    $0.017\pm0.006$\\
    $\Xi_{bc}^{0}\to\Sigma_{c}^{*+}\tau^{-}\bar{\nu}_{\tau} $ &$(
    2.87\pm0.52)\times 10^{-17}$ &
    $(
    4.06\pm0.73)\times 10^{-6}$ &
    $0.004\pm0.001$\\
    $\Xi_{bc}^{+}\to\Sigma_{c}^{*++}l^{-}\bar{\nu}_{l} $ &$(
    2.93\pm0.65)\times 10^{-17}$ &
    $(
    1.09\pm0.24)\times 10^{-5}$ &
    $0.017\pm0.006$\\
    $\Xi_{bc}^{+}\to\Sigma_{c}^{*++}\tau^{-}\bar{\nu}_{\tau} $ &$(
    5.73\pm1.03)\times 10^{-17}$ &
    $(
    2.13\pm0.38)\times 10^{-5}$ &
    $0.004\pm0.001$\\
    $\Xi_{bb}^{-}\to\Sigma_{b}^{*0}l^{-}\bar{\nu}_{l} $ &$(
    7.72\pm1.80)\times 10^{-17}$ &
    $(
    4.34\pm1.01)\times 10^{-5}$ &
    $0.265\pm0.093$\\
    $\Xi_{bb}^{-}\to\Sigma_{b}^{*0}\tau^{-}\bar{\nu}_{\tau} $ &$(
    1.16\pm0.26)\times 10^{-16}$ &
    $(
    6.50\pm1.44)\times 10^{-5}$ &
    $0.082\pm0.026$\\
    $\Xi_{bb}^{0}\to\Sigma_{b}^{*+}l^{-}\bar{\nu}_{l} $ &$(
    1.55\pm0.36)\times 10^{-16}$ &
    $(
    8.74\pm2.04)\times 10^{-5}$ &
    $0.265\pm0.093$\\
    $\Xi_{bb}^{0}\to\Sigma_{b}^{*+}\tau^{-}\bar{\nu}_{\tau} $ &$(
    2.33\pm0.52)\times 10^{-16}$ &
    $(
    1.31\pm0.29)\times 10^{-4}$ &
    $0.083\pm0.026$\\
    \hline\hline
    \end{tabular} 
   \end{table}

\section{Conclusions}
The discovery of $\Xi_{cc}^{++}$ by LHCb collaboration in 2017 has inspired the interest in studying doubly charmed baryons. In this work,
we have investigated the semi-leptonic decays $\Xi_{QQ^{\prime}}\to\Sigma_{Q^{\prime}}^{*}$ using the LCSR approach, which is a continuation of our previous works, extending the spin parity of the final baryon states from $1/2^{+}$ to $3/2^{+}$.
In this study, we present the first LCSR calculation of the form factors of the transition $\Xi_{QQ^{\prime}}\to\Sigma_{Q^{\prime}}^{*}$ with the parallel LCDAs of $\Sigma_{Q^{\prime}}^{*}$.
Among that, the quark-hadron duality has been used to generate the form factors for the decay of $\Xi_{QQ^{\prime}}$ with positive parity.
The numerical results of these form factors are given in Tab.~\ref{Tab:ff_all}, which can be served as inputs for the calculation of non-leptonic decays in the future research.
Our theoretical results for the decay widths and branching ratios of $\Xi_{QQ^{\prime}}\to\Sigma_{Q^{\prime}}^{*}l\nu$ are predicted, which is shown in Tab.~~\ref{Tab:semi_lep}.
The error estimation and theoretical analyses about them are also given in detail.
In addition, a set of SU(3) relations between the decay widths of such processes can be found in Eq.~(\ref{su3symmetry}).
As an estimation, we give the decay widths for $\Omega_{QQ^{\prime}}^{+}\to\Xi_{Q^{\prime}}^{\prime*0}l^{+}\nu$ using SU(3) symmetry.
We hope our investigation can provide valuable suggestions for experimental research about doubly heavy baryons conducted by the LHCb and other experiments and help people to get a comprehensive understanding of the dynamics of baryon decays in the future.
\label{sec:conclusions}

\section{Appendix}
In Eq.~(\ref{eq:parameterization1}), the 3/2-spinor ${u}_{\alpha}$ satisfies $\gamma^{\alpha}{u}_{\alpha}=0$ and $p_{\Sigma^*}^{\alpha}{u}_{\alpha}=0$. All the Lorentz structures ${\cal LS}_{i}^{\alpha\mu}$ multiplying on the form factors $f_{i}$ or $g_{i}$ satisfy $q^{\mu}{\cal LS}_{1,2,4}^{\alpha\mu}=0$. In Ref.~\cite{Zwicky:2013eda,Hiller:2021zth}, there is another parameterization satisfying the relation $q^{\mu}{\cal LS}_{1,2,4}^{\alpha\mu}=0$, which reads as:
\begin{eqnarray}
  &&	\langle{\Sigma_{Q^{\prime}}^{*}}(P^{\prime},S^{\prime}=\frac{3}{2},S_{z}^{\prime})|(V-A)^{\mu}|{\Xi_{QQ^{\prime}q}}(P,S=\frac{1}{2},S_{z})\rangle \nonumber\\
  && =  \bar{u}_{\alpha}(P^{\prime},S_{z}^{\prime})\Big\{\frac{{f}_{1}^{\prime}(q^2)}{M_{1}}({\slashed q}g^{\alpha\mu}-q^{\alpha}\gamma^{\mu})+\frac{{f}_{2}^{\prime}(q^2)}{M_{1}^2}[{q^{\alpha}p_{\Sigma^*}^{\mu}}-\frac{1}{2}(M_{1}^2-M_{2}^{2}-q^2)g^{\alpha\mu}]\nonumber\\
  &&\quad\quad\quad\quad\quad\quad
  +{f}_{3}^{\prime}(q^2)\frac{q^{\alpha}q^{\mu}}{q^2}+\frac{{f}_{4}^{\prime}(q^{2})}{M_{1}^2}(q^2g^{\alpha\mu}-{q^{\alpha}q^{\mu}})\Big\}\gamma_{5}u(P,S_{z})\nonumber\\
  & &\quad- \bar{u}_{\alpha}(P^{\prime},S_{z}^{\prime})\Big\{\frac{{g}_{1}^{\prime}(q^2)}{M_{1}}(q^{\alpha}\gamma^{\mu}-{\slashed q}g^{\alpha\mu})+\frac{{g}_{2}^{\prime}(q^2)}{M_{1}^2}[{q^{\alpha}p_{\Sigma^*}^{\mu}}-\frac{1}{2}(M_{1}^2-M_{2}^{2}-q^2)g^{\alpha\mu}]\nonumber\\
  &&\quad\quad\quad\quad\quad\quad
  +{g}_{3}^{\prime}(q^2)\frac{q^{\alpha}q^{\mu}}{q^2}+\frac{{g}_{4}^{\prime}(q^{2})}{M_{1}^2}(q^2g^{\alpha\mu}-{q^{\alpha}q^{\mu}})\}u(P,S_{z}).\label{eq:parameterization11}
\end{eqnarray}
The form factors $f_{i}^{\prime}$ and $g_{i}^{\prime}$ in Eq.~(\ref{eq:parameterization11}) have the following relationships with the ones in Eq.~(\ref{eq:parameterization1}),
\begin{eqnarray}
  &&f^{\prime}_{1}=-f_{1},~f_{2}^{\prime}=-2f_{2},~f_{3}^{\prime}=\frac{M_{1}^2-M_{2}^{2}}{M_{1}^2q^2}f_{3},~
  f_{4}^{\prime}= \frac{M_{1}^2}{q^2}(\frac{M_{1}+M_{2}}{M_{1}}f_{1}-\frac{M_{1}^2-M_{2}^{2}-q^2}{M_{1}^2}f_{2}+f_{4}) \nonumber\\
  &&	g^{\prime}_{1}=g_{1},~ g_{2}^{\prime}=-2g_{2},~g_{3}^{\prime}=\frac{{M_{1}^2-M_{2}^{2}}}{{M_{1}^2q^2}}g_{3},~
  g_{4}= \frac{M_{1}^2}{q^2}(\frac{M_{1}-M_{2}}{M_{1}}g_{1}-\frac{M_{1}^2-M_{2}^{2}-q^2}{M_{1}^2}g_{2}+g_{4}).
  \nonumber
 \label{eq:relations1}
\end{eqnarray}
These form factors also have relationship with the form factors $F_{i}$ and $G_{i}$ in Eq.~(\ref{eq:parameterization2}). Here we take the $f_{i}^{\prime}$ as an example, 
\begin{align}
  & {f}_{1}^{\prime}(q^2)=M_{1}F_{1}(q^2),\quad
 {f}_{2}^{\prime}(q^2)={M_{1}^2}[F_{2}(q^2)
 +F_{3}(q^2)],\label{eq:f12}\quad\\
 & {f}_{4}^{\prime}(q^2)=\frac{M_{1}^{2}}{2q^2}[-2(M_{1}+M_{2})]F_{1}(q^2)+(M_{1}^2-M_{2}^{2}-q^2)[F_{2}(q^2)
 +F_{3}(q^2)],\label{eq:f14}\\
  & {f}_{3}^{\prime}(q^2)=\frac{M_{1}^{2}}{M_{1}^2-M_{2}^{2}}
 \Big[F_{1}(q^2){(-M_{1}-M_{2})}+F_{4}(q^2)\Big]
 +\frac{M_{1}^2}{2}\Big[F_{2}(q^2)
 +F_{3}(q^2)\Big]\nonumber \\
  & \qquad\qquad\quad+\frac{1}{2}\frac{q^{2}M_{1}^2}{M_{1}^2-M_{2}^{2}}
 \Big[F_{2}(q^2)
 -F_{3}(q^2)\Big],\label{eq:f13}
\end{align}
and the $g_{i}$ have similar relationship with $G_{i}$. Using $f_{i}^{\prime}$ and $g_{i}^{\prime}$ we can write the helicity amplitude as
\begin{eqnarray}
  &&H_{{3/2},1}^{V} =  -i\sqrt{Q_{-}}\big[\frac{M_{1}+M_{2}}{M_{1}}f_{1}^{\prime}+\frac{q^2-(M_{1}^2-M_{2}^{2})}{2M_{1}^2}f_{2}^{\prime}+\frac{q^2}{M_{1}^2}f_{4}^{\prime}\big],\nonumber\\
  &&H_{{3/2},1}^{A} =  i\sqrt{Q_{+}}\big[-\frac{M_{1}-M_{2}}{M_{1}}g_{1}^{\prime}+\frac{q^2-(M_{1}^2-M_{2}^{2})}{2M_{1}^2}g_{2}^{\prime}+\frac{q^2}{M_{1}^2}g_{4}^{\prime}\big],\nonumber\\
  &&H_{\frac{1}{2},1}^{V} =  i\sqrt{\frac{Q_{-}}{3}}\big[\frac{q^2-M_{1}(M_{1}+M_{2})}{M_{1}M_{2}}f_{1}^{\prime}-\frac{q^2-(M_{1}^2-M_{2}^{2})}{2M_{1}^{2}}f_{2}^{\prime}+\frac{q^2}{M_{1}^{2}}f_{4}^{\prime}\big],\nonumber \\
  &&H_{\frac{1}{2},1}^{A} =  i\sqrt{\frac{Q_{+}}{3}}\big[\frac{q^2-M_{1}(M_{1}-M_{2})}{M_{1}M_{2}}g_{1}^{\prime}+\frac{q^2-(M_{1}^2-M_{2}^{2})}{2M_{1}^{2}}g_{2}^{\prime}+\frac{q^2}{M_{1}^{2}}g_{4}^{\prime}\big],\nonumber \\
  &&H_{{1/2},0}^{V}  = -i\frac{\sqrt{2q^{2}Q_{-}}}{\sqrt{3}M}\big[f_{1}^{\prime}+\frac{M_{2}}{M_{1}}f_{2}^{\prime}+\frac{q^2-(M_{1}^{2}-M_{2}^{2})}{2M_{1}M_{2}}f_{4}^{\prime}\big],\; H_{-1/2,t}^{V}=i\frac{\sqrt{Q_{+}}}{\sqrt{6q^{2}}}\frac{Q_{-}}{M_{2}}f_{3}^{\prime},\nonumber \\
  &&H_{{1/2},0}^{A}  =  i\frac{\sqrt{2q^{2}Q_{+}}}{\sqrt{3}M}\big[g_{1}^{\prime}+\frac{M_{2}}{M_{1}}g_{2}^{\prime}+\frac{q^2-(M_{1}^{2}-M_{2}^{2})}{2M_{1}M_{2}}g_{4}^{\prime}\big],\; H_{-1/2,t}^{A}=-i\frac{\sqrt{Q_{-}}}{\sqrt{6q^{2}}}\frac{Q_{+}}{M_{2}}g_{3}^{\prime}.
  \label{eq:helicityamp11}  
\end{eqnarray}
As argued by Ref.~\cite{Zwicky:2013eda,Hiller:2021zth}, at the endpoint $k=\frac{\sqrt{\lambda(M_{1}^2,M_{2}^2,q^2)}}{2M_{1}}\to 0$ with $\lambda(a,b,c)=a^2+b^2+c^2$,  
the above helicity amplitudes should satisfy $H_{0,1/2}:H_{1,1/2}:H_{1,3/2}= 1:-\sqrt{1/2}:-\sqrt{3/2}$. Using the relationship between $f_{i}^{\prime},\ g_{i}^{\prime}$ and $f_{i},\ g_{i}$, We can prove that our helicity amplitudes given in Eq.~(\ref{eq:helicityampli}) also satisfy $H_{0,1/2}:H_{1,1/2}:H_{1,3/2}= 1:-\sqrt{1/2}:-\sqrt{3/2}$.

\section*{Acknowledgements}
The authors are very grateful to Prof. Wei Wang for useful discussions. This work is supported in part by the 13th Sailing Plan No.102521101,
and the youth Foundation JN210003, of China University of mining and technology.


\begin{thebibliography}{10}
\bibitem{Gell-Mann:1964ewy}
M.~Gell-Mann,
Phys. Lett. \textbf{8}, 214-215 (1964)
doi:10.1016/S0031-9163(64)92001-3

\bibitem{Zweig:1964jf}
G.~Zweig,
CERN-TH-412.

\bibitem{DeRujula:1975qlm}
A.~De Rujula, H.~Georgi and S.~L.~Glashow,
Phys. Rev. D \textbf{12}, 147-162 (1975)
doi:10.1103/PhysRevD.12.147

\bibitem{Jaffe:1975us}
R.~L.~Jaffe and J.~E.~Kiskis,
Phys. Rev. D \textbf{13}, 1355 (1976)
doi:10.1103/PhysRevD.13.1355

\bibitem{Ponce:1978gk}
W.~Ponce,
Phys. Rev. D \textbf{19}, 2197 (1979)
doi:10.1103/PhysRevD.19.2197

\bibitem{Fleck:1988vm}
S.~Fleck, B.~Silvestre-Brac and J.~M.~Richard,
Phys. Rev. D \textbf{38}, 1519-1529 (1988)
doi:10.1103/PhysRevD.38.1519

\bibitem{Ebert:2002ig}
D.~Ebert, R.~N.~Faustov, V.~O.~Galkin and A.~P.~Martynenko,
Phys. Rev. D \textbf{66}, 014008 (2002)
doi:10.1103/PhysRevD.66.014008
[arXiv:hep-ph/0201217 [hep-ph]].

\bibitem{Roberts:2007ni}
W.~Roberts and M.~Pervin,
Int. J. Mod. Phys. A \textbf{23}, 2817-2860 (2008)
doi:10.1142/S0217751X08041219
[arXiv:0711.2492 [nucl-th]].

\bibitem{Karliner:2014gca}
M.~Karliner and J.~L.~Rosner,
Phys. Rev. D \textbf{90}, no.9, 094007 (2014)
doi:10.1103/PhysRevD.90.094007
[arXiv:1408.5877 [hep-ph]].

\bibitem{He:2004px}
D.~H.~He, K.~Qian, Y.~B.~Ding, X.~Q.~Li and P.~N.~Shen,
Phys. Rev. D \textbf{70}, 094004 (2004)
doi:10.1103/PhysRevD.70.094004
[arXiv:hep-ph/0403301 [hep-ph]].

\bibitem{Bagan:1992za}
E.~Bagan, M.~Chabab and S.~Narison,
Phys. Lett. B \textbf{306}, 350-356 (1993)
doi:10.1016/0370-2693(93)90090-5

\bibitem{Valcarce:2008dr}
A.~Valcarce, H.~Garcilazo and J.~Vijande,
Eur. Phys. J. A \textbf{37}, 217-225 (2008)
doi:10.1140/epja/i2008-10616-4
[arXiv:0807.2973 [hep-ph]].

\bibitem{Wang:2010hs}
Z.~G.~Wang,
Eur. Phys. J. A \textbf{45}, 267-274 (2010)
doi:10.1140/epja/i2010-11004-3
[arXiv:1001.4693 [hep-ph]].

\bibitem{Kiselev:2001fw}
V.~V.~Kiselev and A.~K.~Likhoded,
Phys. Usp. \textbf{45}, 455-506 (2002)
doi:10.1070/PU2002v045n05ABEH000958
[arXiv:hep-ph/0103169 [hep-ph]].

\bibitem{Zhang:2008rt}
J.~R.~Zhang and M.~Q.~Huang,
Phys. Rev. D \textbf{78}, 094007 (2008)
doi:10.1103/PhysRevD.78.094007
[arXiv:0810.5396 [hep-ph]].

\bibitem{Aliev:2012ru}
T.~M.~Aliev, K.~Azizi and M.~Savci,
Nucl. Phys. A \textbf{895}, 59-70 (2012)
doi:10.1016/j.nuclphysa.2012.09.009
[arXiv:1205.2873 [hep-ph]].

\bibitem{Richard:2005jz}
J.~M.~Richard and F.~Stancu,
Bled Workshops Phys. \textbf{6}, no.1, 25-31 (2005)
[arXiv:hep-ph/0511043 [hep-ph]].

\bibitem{Lewis:2001iz}
R.~Lewis, N.~Mathur and R.~M.~Woloshyn,
Phys. Rev. D \textbf{64}, 094509 (2001)
doi:10.1103/PhysRevD.64.094509
[arXiv:hep-ph/0107037 [hep-ph]].

\bibitem{Flynn:2003vz}
J.~M.~Flynn \textit{et al.} [UKQCD],
JHEP \textbf{07}, 066 (2003)
doi:10.1088/1126-6708/2003/07/066
[arXiv:hep-lat/0307025 [hep-lat]].

\bibitem{Liu:2009jc}
L.~Liu, H.~W.~Lin, K.~Orginos and A.~Walker-Loud,
Phys. Rev. D \textbf{81}, 094505 (2010)
doi:10.1103/PhysRevD.81.094505
[arXiv:0909.3294 [hep-lat]].

\bibitem{SELEX:2002wqn}
M.~Mattson \textit{et al.} [SELEX],
Phys. Rev. Lett. \textbf{89}, 112001 (2002)
doi:10.1103/PhysRevLett.89.112001
[arXiv:hep-ex/0208014 [hep-ex]].

\bibitem{Ratti:2003ez}
S.~P.~Ratti,
Nucl. Phys. B Proc. Suppl. \textbf{115}, 33-36 (2003)
doi:10.1016/S0920-5632(02)01948-5

\bibitem{BaBar:2006bab}
B.~Aubert \textit{et al.} [BaBar],
Phys. Rev. D \textbf{74}, 011103 (2006)
doi:10.1103/PhysRevD.74.011103
[arXiv:hep-ex/0605075 [hep-ex]].

\bibitem{Belle:2006edu}
R.~Chistov \textit{et al.} [Belle],
Phys. Rev. Lett. \textbf{97}, 162001 (2006)
doi:10.1103/PhysRevLett.97.162001
[arXiv:hep-ex/0606051 [hep-ex]].

\bibitem{LHCb:2017iph}
R.~Aaij \textit{et al.} [LHCb],
Phys. Rev. Lett. \textbf{119}, no.11, 112001 (2017)
doi:10.1103/PhysRevLett.119.112001
[arXiv:1707.01621 [hep-ex]].

\bibitem{LHCb:2018pcs}
R.~Aaij \textit{et al.} [LHCb],
Phys. Rev. Lett. \textbf{121}, no.16, 162002 (2018)
doi:10.1103/PhysRevLett.121.162002
[arXiv:1807.01919 [hep-ex]].

\bibitem{LHCb-PAPER-2018-019}
R.~Aaij \textit{et al.} [LHCb],
Phys. Rev. Lett. \textbf{121}, no.5, 052002 (2018)
doi:10.1103/PhysRevLett.121.052002
[arXiv:1806.02744 [hep-ex]].

\bibitem{LHCb-PAPER-2019-035}
R.~Aaij \textit{et al.} [LHCb],
Chin. Phys. C \textbf{44}, no.2, 022001 (2020)
doi:10.1088/1674-1137/44/2/022001
[arXiv:1910.11316 [hep-ex]].

\bibitem{LHCb-PAPER-2019-037}
R.~Aaij \textit{et al.} [LHCb],
JHEP \textbf{02}, 049 (2020)
doi:10.1007/JHEP02(2020)049
[arXiv:1911.08594 [hep-ex]].

\bibitem{Shi:2017dto}
Y.~J.~Shi, W.~Wang, Y.~Xing and J.~Xu,
Eur. Phys. J. C \textbf{78}, no.1, 56 (2018)
doi:10.1140/epjc/s10052-018-5532-7
[arXiv:1712.03830 [hep-ph]].

\bibitem{Wang:2021uzi}
R.~M.~Wang, Y.~G.~Xu, C.~Hua and X.~D.~Cheng,
Phys. Rev. D \textbf{103}, no.1, 013007 (2021)
doi:10.1103/PhysRevD.103.013007
[arXiv:2101.02421 [hep-ph]].

\bibitem{Li:2021rfj}
D.~M.~Li, X.~R.~Zhang, Y.~Xing and J.~Xu,
Eur. Phys. J. Plus \textbf{136}, no.7, 772 (2021)
doi:10.1140/epjp/s13360-021-01757-6
[arXiv:2101.12574 [hep-ph]].

\bibitem{Shi:2019hbf}
Y.~J.~Shi, W.~Wang and Z.~X.~Zhao,
Eur. Phys. J. C \textbf{80}, no.6, 568 (2020)
doi:10.1140/epjc/s10052-020-8096-2
[arXiv:1902.01092 [hep-ph]].

\bibitem{Zhao:2020mod}
Z.~X.~Zhao, R.~H.~Li, Y.~L.~Shen, Y.~J.~Shi and Y.~S.~Yang,
Eur. Phys. J. C \textbf{80}, no.12, 1181 (2020)
doi:10.1140/epjc/s10052-020-08767-1
[arXiv:2010.07150 [hep-ph]].

\bibitem{Yu:2017zst}
F.~S.~Yu, H.~Y.~Jiang, R.~H.~Li, C.~D.~L\"u, W.~Wang and Z.~X.~Zhao,
Chin. Phys. C \textbf{42}, no.5, 051001 (2018)
doi:10.1088/1674-1137/42/5/051001
[arXiv:1703.09086 [hep-ph]].

\bibitem{Wang:2017mqp}
W.~Wang, F.~S.~Yu and Z.~X.~Zhao,
Eur. Phys. J. C \textbf{77}, no.11, 781 (2017)
doi:10.1140/epjc/s10052-017-5360-1
[arXiv:1707.02834 [hep-ph]].

\bibitem{Hu:2020mxk}
X.~H.~Hu, R.~H.~Li and Z.~P.~Xing,
Eur. Phys. J. C \textbf{80}, no.4, 320 (2020)
doi:10.1140/epjc/s10052-020-7851-8
[arXiv:2001.06375 [hep-ph]].

\bibitem{Guo:2005qa}
P.~Guo, H.~W.~Ke, X.~Q.~Li, C.~D.~Lu and Y.~M.~Wang,
Phys. Rev. D \textbf{75}, 054017 (2007)
doi:10.1103/PhysRevD.75.054017
[arXiv:hep-ph/0501058 [hep-ph]].

\bibitem{Lu:2009cm}
C.~D.~Lu, Y.~M.~Wang, H.~Zou, A.~Ali and G.~Kramer,
Phys. Rev. D \textbf{80}, 034011 (2009)
doi:10.1103/PhysRevD.80.034011
[arXiv:0906.1479 [hep-ph]].

\bibitem{Wang:2008sm}
Y.~m.~Wang, Y.~Li and C.~D.~Lu,
Eur. Phys. J. C \textbf{59}, 861-882 (2009)
doi:10.1140/epjc/s10052-008-0846-5
[arXiv:0804.0648 [hep-ph]].

\bibitem{Wang:2009hra}
Y.~M.~Wang, Y.~L.~Shen and C.~D.~Lu,
Phys. Rev. D \textbf{80}, 074012 (2009)
doi:10.1103/PhysRevD.80.074012
[arXiv:0907.4008 [hep-ph]].

\bibitem{Shi:2019fph}
Y.~J.~Shi, Y.~Xing and Z.~X.~Zhao,
Eur. Phys. J. C \textbf{79}, no.6, 501 (2019)
doi:10.1140/epjc/s10052-019-7014-y
[arXiv:1903.03921 [hep-ph]].

\bibitem{Hu:2019bqj}
X.~H.~Hu and Y.~J.~Shi,
Eur. Phys. J. C \textbf{80}, no.1, 56 (2020)
doi:10.1140/epjc/s10052-020-7635-1
[arXiv:1910.07909 [hep-ph]].

\bibitem{Ali:2012pn}
A.~Ali, C.~Hambrock, A.~Y.~Parkhomenko and W.~Wang,
Eur. Phys. J. C \textbf{73}, no.2, 2302 (2013)
doi:10.1140/epjc/s10052-013-2302-4
[arXiv:1212.3280 [hep-ph]].

\bibitem{Shah:2016vmd}
Z.~Shah, K.~Thakkar and A.~K.~Rai,
Eur. Phys. J. C \textbf{76}, no.10, 530 (2016)
doi:10.1140/epjc/s10052-016-4379-z
[arXiv:1609.03030 [hep-ph]].

\bibitem{Shah:2017liu}
Z.~Shah and A.~K.~Rai,
Eur. Phys. J. C \textbf{77}, no.2, 129 (2017)
doi:10.1140/epjc/s10052-017-4688-x
[arXiv:1702.02726 [hep-ph]].

\bibitem{Groote:1997yr}
S.~Groote, J.~G.~Korner and O.~I.~Yakovlev,
Phys. Rev. D \textbf{56}, 3943-3954 (1997)
doi:10.1103/PhysRevD.56.3943
[arXiv:hep-ph/9705447 [hep-ph]].

\bibitem{Hu:2017dzi}
X.~H.~Hu, Y.~L.~Shen, W.~Wang and Z.~X.~Zhao,
Chin. Phys. C \textbf{42}, no.12, 123102 (2018)
doi:10.1088/1674-1137/42/12/123102
[arXiv:1711.10289 [hep-ph]].

\bibitem{Aaij:2017ueg}
R.~Aaij \textit{et al.} [LHCb],
Phys. Rev. Lett. \textbf{119}, no.11, 112001 (2017)
doi:10.1103/PhysRevLett.119.112001
[arXiv:1707.01621 [hep-ex]].

\bibitem{Aaij:2018wzf}
R.~Aaij \textit{et al.} [LHCb],
Phys. Rev. Lett. \textbf{121}, no.5, 052002 (2018)
doi:10.1103/PhysRevLett.121.052002
[arXiv:1806.02744 [hep-ex]].

\bibitem{Cheng:2018mwu}
H.~Y.~Cheng and Y.~L.~Shi,
Phys. Rev. D \textbf{98}, no.11, 113005 (2018)
doi:10.1103/PhysRevD.98.113005
[arXiv:1809.08102 [hep-ph]].

\bibitem{Brown:2014ena}
Z.~S.~Brown, W.~Detmold, S.~Meinel and K.~Orginos,
Phys. Rev. D \textbf{90}, no.9, 094507 (2014)
doi:10.1103/PhysRevD.90.094507
[arXiv:1409.0497 [hep-lat]].

\bibitem{Olive:2016xmw}
C.~Patrignani \textit{et al.} [Particle Data Group],
Chin. Phys. C \textbf{40}, no.10, 100001 (2016)
doi:10.1088/1674-1137/40/10/100001

\bibitem{Tanabashi:2018oca}
M.~Tanabashi \textit{et al.} [Particle Data Group],
Phys. Rev. D \textbf{98}, no.3, 030001 (2018)
doi:10.1103/PhysRevD.98.030001

\bibitem{Wang:2017azm}
W.~Wang, Z.~P.~Xing and J.~Xu,
Eur. Phys. J. C \textbf{77}, no.11, 800 (2017)
doi:10.1140/epjc/s10052-017-5363-y
[arXiv:1707.06570 [hep-ph]].

\bibitem{Zwicky:2013eda}
R.~Zwicky,
Nucl. Phys. B \textbf{975}, 115673 (2022)
doi:10.1016/j.nuclphysb.2022.115673
[arXiv:1309.7802 [hep-ph]].

\bibitem{Hiller:2021zth}
G.~Hiller and R.~Zwicky,
JHEP \textbf{11}, 073 (2021)
doi:10.1007/JHEP11(2021)073
[arXiv:2107.12993 [hep-ph]].
\end{thebibliography}
\end{document}